\documentclass[11pt]{amsart}
\usepackage{amsbsy,amssymb,amsmath,amsthm,amscd,amsfonts,latexsym,amstext,delarray,
amsmath,graphicx, url} 
\usepackage[margin=1in]{geometry}
\usepackage{color}
\input xypic

\usepackage[all]{xy}
\usepackage{setspace}

\usepackage{amsaddr}

\newtheorem{thm}{Theorem}[section]
\newtheorem{prop}[thm]{Proposition}

\newtheorem{lem}[thm]{Lemma}

\newtheorem{defn}[thm]{Definition}
\newtheorem{rem}[thm]{Remark}
\newtheorem{ex}[thm]{Example}

\numberwithin{equation}{section}

\def\F{{\mathbb F}}
\def\Q{{\mathbb Q}}
\def\Z{{\mathbb Z}}
\def\N{{\mathbb N}}
\def\R{{\mathbb R}}
\def\C{{\mathbb C}}
\def\K{{\mathbb K}}

\def\P{{\mathbb P}}
\def\A{{\mathbb A}}
\def\GL{{\rm GL}}

\def\cD{{\mathcal D}}

\def\cH{{\mathcal H}}
\def\cI{{\mathcal I}}

\def\cL{{\mathcal L}}
\def\cM{{\mathcal M}}

\def\cO{{\mathcal O}}
\def\cP{{\mathcal P}}

\def\cS{{\mathcal S}}
\def\cT{{\mathcal T}}

\def\cV{{\mathcal V}}

\def\bA{{\mathbb A}}

\def\bG{{\mathbb G}}

\def\bP{{\mathbb P}}

\def\bT{{\mathbb T}}

\def\fm{{\mathfrak m}}

\def\PGL{{\rm PGL}}

\title{Adelic Models of Percolation}
\author{Matilde Marcolli}
\date{2025}

\address{Department of Mathematics and Department of Computing and Mathematical Sciences, 
California Institute of Technology, Pasadena, CA 91125, USA}
\email{matilde@caltech.edu}

\begin{document}

\maketitle
\makeatletter
\vspace{-2em} 
{\centering\enddoc@text}
\let\enddoc@text\empty 
\makeatother

\begin{abstract}
Models of long range percolation on lattices and on hierarchical lattices are related through the use of
three intermediate geometries: a $1$-parameter deformation based on the power mean function, relating
lattice percolation to a percolation model governed by the toric volume form; the adelic product formula for
a function field, relating the hierarchical lattice model to an adelic percolation model; and the adelic product 
formula for number fields that relates the toric percolation model on the lattice given by the ring of
integers in the Minkowski embedding to another adelic percolation model. 
\end{abstract}

\begin{center}
This paper was written for the occasion of the Conference in Memory of Yuri Manin,
Max Planck Institute for Mathematics, Bonn, August 11--15, 2025
\end{center}

\section{Introduction}

In 1989, Manin published {\em Reflections on Arithmetical Physics}, \cite{ManArPhys}, where he predicted
an increasingly prominent role for number theory within theoretical physics. At that time, early results in
String Theory, including Manin's work on the partition function of the Polyakov string (\cite{BeMa}, \cite{ManPolya}), 
showed that arithmetic concepts like Faltings' height had physical relevance. Since then, there has been 
a wealth of examples
in which this prediction came true: Grothendieck motives and periods, various classes of modular forms, 
and $p$-adic geometry have all found direct applications to physics models, as did many other 
number theoretic objects. Ideas from physics have, in turn, suggested new strategies in classical
number theory problems.

One of the most interesting ideas that Manin articulated in \cite{ManArPhys} revolves around the
adelic product formula involving the $p$-adic non-Archimedean norms and the Archimedean absolute
value of rational numbers through the simple relation 
\begin{equation}\label{prodpadicnormsQ}
\prod_p |x|_p = | x |_\infty^{-1} \, ,  \ \ \ \  \forall x\in \Q\, , x\neq 0\, . 
\end{equation}
This formula is seen as the prototype example illustrating a general strategy, aimed at replacing
concepts usually formulated in terms of the real numbers, where the Archimedean absolute
value lives, by an adelic counterpart, where the product over primes of the non-Archimedean norms
is defined. Using the fact that, generally, these adelic objects turn out to have simpler properties makes
it then possible to establish results that can be transferred back to the real Archimedean place by this type of adelic
formula. He argued that such a strategy should be especially beneficial in physics problems, where
he postulated a form of ``complementarity" between our usual formulation of the fundamental models
of physics in terms of real variables and an equivalent adelic counterpart. 

In our joint work, we first explored this notion in the context of the holographic AdS/CFT correspondence
in string theory in \cite{ManMarHolo}, a set of ideas that I more recently returned to 
in work with students and other collaborators, \cite{HMSS}, \cite{HMPS}.

In this paper, I would like to introduce a different context in which the same procedure of recasting
a real variables formulation of a physical problem in terms of Manin's idea of an ``adelic shadow"
can be applied: the long-range percolation models in statistical physics.

I will show how one can use an adelic formulation to establish a direct geometric relation
between the long-range percolation models on ordinary lattices and on hierarchical
lattices. 

\subsection{Long-range percolation models and their relation}\label{IntroPercRelSec} 

In this paper we consider two different types of long range percolation models. The first is 
long range percolation on a lattice $\Lambda \subset \R^d$. In addition to the case 
$\Lambda=\Z^d$ we will consider lattices that are rings of integers $\Lambda=\cO_\K$ of number fields $\K$,
viewed as embedded in a Euclidean space via the Minkowski embedding determined
by the Archimedean places of the number field. 

The long range percolation model on a lattice
is a random graph with $\Lambda$ as set of vertices, where a pair $x\neq y \in \Lambda$
is connected by an edge with probability $1-\exp(-\beta \| x-y \|^{-d-\alpha})$ for a 
fixed parameter $\alpha >0$ and varying inverse-temperature parameter $\beta \geq 0$.
A main question in such a percolation model is the geometry of clusters (connected components) 
or the resulting random graph and changes occurring at phase transitions  
$\beta_c=\inf \{ \beta \geq 0\,|\,  \bP_\beta(\text{an infinite cluster exists}) > 0 \}$.
In particular one is interested in questions such as the behavior at criticality of the
2-point function, given by the connection probability
$\bP_{\beta_c} (x \leftrightarrow y) \sim \| x-y \|^{-d+2-\eta}$ for $\| x-y \| \to \infty$, the
critical exponent $\eta$ and its relations to other critical exponents such
as the one ruling the volume tail behavior $\bP_{\beta_c}(|K|\geq n)\sim n^{-1/\delta}$,
and understanding the dependence of the model behavior  and the critical exponents on the parameter $\alpha$. 
Similarly, one is interested in the behavior of
the n-point functions given by the probability $\bP_\beta (A)$ that all $x_i\in A$ are in the same cluster 
for a set  $A=\{ x_1, \ldots, x_n \}$ of $n$ vertices. 
For a general overview of various aspects of these percolation models on lattices and the main
questions about them see for instance \cite{HeyHof}. 

The other class of long range percolation models we consider are those on
hierarchical lattices. Here instead of a lattice $\Lambda$ one considers an abelian
group of the form $\cH^d_L :=\oplus_{i=1}^\infty \bT^d_L$, where $\bT^d_L:= (\Z/L\Z)^d$ 
for some integers $d\geq 1$ and $L\geq 2$. and connection probabilities 
$1- \exp(-\beta\| x-y \|_{\cH^d_L}^{-d -\alpha})$,  with the ultrametric
$\| x-y \|_{\cH^d_L}= L^{h(x,y)}$ for $x\neq y$, with $h(x,y)= \max\{ i\geq 1 \,|\, x_i\neq y_i \}$.
Unlike ordinary lattices, the hierarchical lattice has a self-similar fractal-like geometry, \cite{DaGo}.

Recent work of Hutchcroft 
shows that the model on hierarchical lattices can sometimes be used to obtain
results on the behavior of the long-range percolation on ordinary lattices,  see for
instance \cite{Hu1}, \cite{Hu2}, \cite{Hu3}, and in particular estimates on the 
critical exponents, see for instance \cite{Hu1}. 

In this paper we consider the abstract geometric question of understanding
the relation between  long range percolation on hierarchical lattices and on ordinary lattices.
We approach the question of their relation from a different perspective, based on
the idea that real variables physical systems have adelic counterparts, and we 
describe a direct relation between the long range effects of these two types of
models through an intermediate arithmetic picture model based on global fields 
(function fields and number fields) and the respective adeles rings. 
Our goal here is to present this arithmetic relation. 

A typical way in which relations between two different geometries are established, in the
context of algebraic or arithmetic geometry, is by realizing them as different special
fibers of some fibration. This same idea underlies, for instance, Arakelov geometry
where an arithmetic curve is seen as a fibration over the Archimedean and 
non-Archimedean places of a number field, see for instance the discussion in
\cite{ManHyp} and \cite{ManMarHolo} on the fiber at infinity of Arakelov geometry.

We follow a similar reasoning here and we realize both the long-range percolation models
on ordinary lattices and those on hierarchical lattices as ``fibers at infinity" of adelic
constructions, in number fields and function fields, respectively, where the fibers over
the individual finite places, give in both cases equivalent systems. In the case of
percolation on ordinary lattices an intermediate step is necessary in the construction
we present, where the usual percolation model is realized as a special value at $t=2$ of
a $1$-parameter deformation determined by the power mean function, whose value at $t=0$
is the fiber at infinity (the contribution of the Archimedean places) of the adelic model.
We summarize these construction that establish the relation between the two types
of percolation models in terms of a diagram of the form:
\begin{equation}\label{bigdiagram}
\xymatrix{  \cP_{\bA_{\F_q(C),\infty}}(\cH^1_q) \ar[d]\ar[r]  & \cP_{\bA_{\F_q(C),{\rm fin}}}(\cH^1_q) \ar[l] \ar[d]  & \cP_{\bA_{\K,{\rm fin}}}(\cO_\K) \ar[d]   \ar[r] & \cP_{\bA_{\K,\infty}}(\cO_\K) \ar[d]  \ar[l]    & \\
 \cP_{\rm hier}(\cH_q^1) \ar[u]  & \cP_{\bA_{\F_q(C),x\neq \infty}}(\cH_q^1) \ar[r] & \cP_{\bA_{\K,{\rm fin}}, \nu}(\cO_\K)  \ar[l]  & \cP_\bT(\cO_\K) \ar[u] \ar[r]^{\cM_t} & \cP_{\rm latt}(\cO_\K)  \ar[l] }
\end{equation}

On the left side of the diagram,
$\F_q(C)$ is a function field and $\cP_{\bA_{\F_q(C),{\rm fin}}}(\cH^1_q)$ is its adelic percolation model
(introduced in \S \ref{FFsec}) with $\cP_{\bA_{\F_q(C),\infty}}(\cH_q^1)$ its component at a point at $\infty$ 
of the algebraic curve $C$ and $\cP_{\bA_{\F_q(C),x\neq \infty}}(\cH_q^1)$ the components at the
finite places, which form the finite adeles part $\cP_{\bA_{\F_q(C),{\rm fin}}}(\cH^1_q)$. The
component at the point at infinity is  the hierarchical lattice percolation model $\cP(\cH_q^1)$, while
the adelic product formula establishes an equivalence in long range behavior between this component
at the point at infinity and the finite adeles part $\cP_{\bA_{\F_q(C),{\rm fin}}}(\cH^1_q)$.

On the right side of the diagram, $\K$ is a number field with
$\cP_{\bA_{\K,{\rm fin}}}(\cO_\K)$ its adelic percolation model over (a region of) the ring of integers $\cO_\K$,
with $\cP_{\bA_{\K,{\rm fin}}}(\cO_\K)$ and $\cP_{\bA_{\K,\infty}}(\cO_\K)$, respectively, the controbutions
of the non-Archimedean places and the Archimedean places, again related by the adelic product formula.
In this case, each non-Archimedean component $\cP_{\bA_{\K,{\rm fin}}, \nu}(\cO_\K)$ 
at one of the places $\nu\in \bA_{\K,{\rm fin}}$ is of the same type as the individual components
and $\cP_{\bA_{\F_q(C),x\neq \infty}}(\cH_q^1)$, while the Archimedean part $\cP_{\bA_{\K,\infty}}(\cO_\K)$
is equivalent to what we call toric percolation model on a lattice (introduced in \S \ref{TorPMSec}),
which in turn is related by the power mean deformation $\cM_t$ to the ordinary lattice
percolation $\cP_{\rm latt}(\cO_\K)$. 

We explain the details of the various parts of this diagram through the coming sections. The more
precise statement of the result illustrated in diagram \eqref{bigdiagram} is given in Theorem~\ref{thmdiagr}.

\section{A family of percolation models on a lattice} \label{PMtSec}

In this section we present a one-parameter family of long-range percolation models on a lattice,
which interpolate and generalize the long-range   percolation on $\Z^d$. The reason
for introducing this range of models will become clear in the following section where we 
geometrically compare them to long-range percolation models on hierarchical lattices, using
non-archimedean and adelic geometry.

\subsection{Long-range   percolation on a lattice}\label{LattSec}

Given a lattice $\Lambda \subset \R^N$ (in particular $\Lambda=\Z^N$), consider the
long-range percolation model $\cP(\Lambda)$, namely the random graph 
where two points $x\neq y\in \Lambda$ are joined by an edge
with inclusion probability
\begin{equation}\label{lattJprob}
\bP_{\beta,\alpha}(\{ x,y \}) =  1- e^{-\beta J_\alpha(x-y)} \, , 
\end{equation}
where $\beta >0$ is a thermodynamic parameter (inverse temperature) and $\alpha>0$ is a fixed
parameter, and with the function 
\begin{equation}\label{lattJfunc}
J_\alpha(x-y) = \| x-y \|^{-N - \alpha} \, . 
\end{equation}
with $\| x-y \|$ the Euclidean norm in $\R^N$.

\smallskip

\begin{rem}\label{longrangeregime}{\rm 
Throughout the paper, for a long-range percolation model with inclusion probabilities of the form \eqref{lattJprob} for some
function $J_\alpha(x-y)$ (which may be determined by a norm, as in \eqref{lattJfunc}, or by some
other function), we will refer to the case where $J_\alpha(x-y)$ is {\em sufficiently small} (i.e.~in a
case like \eqref{lattJfunc}, the norm is sufficiently large), meaning the range in which we
can use the approximation
$$ \bP_{\beta,\alpha}(\{ x,y \}) =  1- e^{-\beta J_\alpha(x-y)} \sim  \beta  J_\alpha(x-y) \, .  $$
}\end{rem}

\subsection{The mean power long-range percolation on a lattice} \label{PMSec}

Let $\Delta_N$ denote the $N$-simplex (the set of probability distributions on a set of $N$-points)
$$ \Delta_N=\{ \lambda=(\lambda_i)_{i=1}^N \in \R^N \,|\, \lambda_i\geq 0,\, \sum_{i=1}^N \lambda_i=1 \} . $$

\begin{defn}\label{PowerMean}
The {\em power mean} function $\cM_t(\lambda,x)$, for $t\in \R^*$ and $\lambda=(\lambda_i)_{i=1}^N \in \Delta_N$, 
and for $x=(x_i)_{i=1}^N \in \R^N$ with $x_i \neq 0$ for all $i$, 
is given by
\begin{equation}\label{MtPM}
\cM_t(\lambda,x):= \left( \sum_{i=1}^N \lambda_i \, |x_i |^t \right)^{1/t} \, .
\end{equation} 
This is extended to
\begin{equation}\label{MtPM2}
\begin{array}{lll}
\cM_{-\infty} (\lambda,x) & := & \min_i |x_i |  \,\, ,  \\
\cM_0 (\lambda,x) & := & \prod_i |x_i |^{\lambda_i} \,\, ,  \\
\cM_{+ \infty} (\lambda,x) & := & \max_i | x_i | \, \,  . 
\end{array}
\end{equation}
\end{defn}

The power mean function is well known in information theory, since the R\'enyi entropy ${\rm Ry}_q(\lambda)$
is obtained as ${\rm Ry}_q(\lambda)=\log \cM_{1-q}(\lambda, 1/\lambda)$ for $1/\lambda:=(1/\lambda_i)_{i=1}^N$; 
see \cite{Lein} for a discussion of its properties and use in information. A main property of $\cM_t(\lambda,x)$ that
will be useful in the following is monotonicity (see Theorem~4.2.8 of \cite{Lein}). Namely, the function $\cM_t(\lambda,x)$
satisfies, for all $\lambda\in \Delta_N$ and $x_i \neq 0$, 
\begin{equation}\label{monoMtPM}
\frac{d}{dt} \log \cM_t (\lambda,x) \geq 0  \, .
\end{equation}
This property implies, as a special case, the arithmetic-geometric mean inequality
\begin{equation}\label{geoarmean}
 \frac{1}{N} \sum_{i=1}^N |x_i | \geq \left( \prod_{i=1}^N |x_i | \right)^{1/N} \, , 
\end{equation} 
when comparing $t=0$ and $t=1$. 

\begin{rem}\label{t2Mt}{\rm
We then consider the function
 \begin{equation}\label{meanJfunc}
J_{\alpha,t, \lambda}(x-y) := \cM_t(\lambda,x-y)^{-N - \alpha} \, . 
\end{equation}
For $t=2$ and $\lambda_i=1/N$, for all $i$, the uniform probability in $\Delta_N$, we have
\begin{equation}\label{PMusualJ}
 J_{\alpha, 2, 1/N}(x-y) = N^{(N+\alpha)/2} \, J_\alpha (x-y) \, , 
\end{equation} 
with $J_\alpha (x-y)$ as in \eqref{lattJfunc}.
Thus, we can view $J_{\alpha,t, \lambda}$ as a generalization (incorporating
the probability $\lambda \in \Delta_N$) and deformation (via the parameter $t\in \R$)
of the function $J_\alpha (x-y)$ of \eqref{lattJfunc}.}
\end{rem}

\begin{defn}\label{MPlattperc} {\rm
The power mean long-range percolation model $\cP_{\cM_t}(\Lambda)$ on a lattice $\Lambda \subset \R^N$
is a percolation model where two points $x\neq y\in \Lambda$ are joined by an edge
with inclusion probability
\begin{equation}\label{MPlattJprob}
\bP_{\beta,\alpha,t,\lambda}(x,y) =  1- e^{-\beta J_{\alpha,t, \lambda}(x-y)} \, , 
\end{equation}
with $J_{\alpha,t, \lambda}(x-y)$ as in \eqref{meanJfunc}, and with
parameters $\beta>0$, $\alpha>0$ and $t\in \R$. 
}\end{defn}

\subsection{Toric volume percolation as a special case} \label{TorPMSec}

We have seen in \eqref{PMusualJ} that the usual function $J_\alpha (x-y)$ of
  long-range percolation on a lattice can be seen as a special case of
the power mean long-range percolation on a lattice, for $t=2$ and the
uniform $\lambda=1/N$. We now show another special case that will be
useful in the following. 

\subsubsection{Power mean function with real and complex coordinates} 
First we extend slightly the definition of $\cM_t(\lambda,x)$ to the case where $x$ is replaced by
$(x,z)=(x_1,\ldots, x_r, z_1, \ldots, z_s)$ with $x_i\in \R$ for $i=1,\ldots, r$ and
$z_j \in \C$, for $j=1,\ldots, s$, by simply setting
\begin{equation}\label{MtRealCompl}
\cM_t(\lambda,(x,z)):= \left( \sum_{i=1}^r \lambda_{1,i} \, |x_i |^t + \sum_{j=1}^s \lambda_{2,j} |z_j|^t
+ \sum_{j=1}^s \lambda_{2,j+s} |\bar z_j|^t \right)^{1/t} 
\, ,
\end{equation}
with $|x_i|$ and $|z_j|$ the real and complex absolute value, and 
with $\lambda=(\lambda_{1,1},\ldots,\lambda_{1,r}, \lambda_{2,1},\ldots, \lambda_{2,2s})$ a probability
$\lambda\in \Delta_{r+2s}$. (Note: we write two separate terms in $|z_j|^t$ and $|\bar z_j|^t$ 
because we will later take $z$ and $\bar z$ to be two complex 
conjugate embeddings of a number field in $\C$.)
We extend this, as before, to
\begin{equation}\label{MtPMRealCompl}
\begin{array}{lll}
\cM_{-\infty} (\lambda,(x,z)) & := & \min_{i,j} \{  |x_i |, |z_j| \}  \,\, ,  \\
\cM_0 (\lambda,(x,z))& := & \prod_{i=1}^r |x_i |^{\lambda_{1,i}} \, \prod_{j=1}^s |z_j|^{\lambda_{2,j}} \, \prod_{j=1}^s |\bar z_j|^{\lambda_{2,j+s}} \,\, ,  \\
\cM_{+ \infty} (\lambda,(x,z))& := & \max_{i,j} \{  |x_i |, |z_j| \} \, \,  . 
\end{array}
\end{equation}
Without loss of generality we can assume that $\lambda_{2,j}=\lambda_{2,j+s}$ and write 
\begin{equation}\label{MtRCr2s}
 \cM_t(\lambda,(x,z)) = \left( \sum_{i=1}^r \lambda_{1,i} \, |x_i |^t + 2\sum_{j=1}^s \lambda_{2,j} |z_j|^t \right)^{1/t}  
\end{equation}
\begin{equation}\label{MtRCr2s0} 
 \cM_0 (\lambda,(x,z)) = \prod_{i=1}^r |x_i |^{\lambda_{1,i}} \, \prod_{j=1}^s |z_j|^{2\lambda_{2,j}} \, . 
 \end{equation}
 
 \begin{defn}\label{PMpercRC}{\rm 
 For a lattice $\Lambda \subset \R^r \times \C^s$, the power mean long-range percolation model on $\Lambda$
 has inclusion probability 
 $$ \bP_{\beta,\alpha,t,\lambda}((x,z),(y,w)) =  1- e^{-\beta J_{\alpha,t, \lambda}(x-y, z-w)} $$
 as \eqref{MPlattJprob} with
 \begin{equation}\label{JtRC}
 J_{\alpha,t, \lambda}(x-y, z-w) := \cM_t(\lambda,(x-y,z-w))^{-N - \alpha}
 \end{equation}
 with $\cM_t(\lambda,(x-y,z-w))$ as in  \eqref{MtRealCompl}, \eqref{MtPMRealCompl}. 
 }\end{defn} 

\subsubsection{Toric volume form and toric percolation model}\label{TorVolSec}

The expression $\cM_0 (\lambda,(x,z))$ has a geometric interpretation in terms of toric volume element.

\begin{defn}\label{transLat}
Let $\cD_N$ denote the divisor of coordinate hyperplanes in the affine space $\bA^N$,
with $\cD_N(\R)\subset \bA^N(\R)=\R^N$, where 
\begin{equation}\label{coordD}
\cD_N=\bigcup_{i=1}^N H_i \, , \ \ \ \text{ with } \ \  H_i =\{ x=(x_1,\ldots, x_N)\in \bA^N \,|\, x_i =0 \}\, .
\end{equation}
We say that a lattice $\Lambda \subset \R^N$ is transverse if
\begin{equation}\label{transvLambda}
\Lambda \cap \cD_N(\R) =\{ 0 \}\, .
\end{equation}
\end{defn}

\smallskip

\begin{ex}\label{1DtransLat}{\rm
Every $1$-dimensional lattice, $\Lambda =\lambda \Z$ for some $\lambda>0$,
is transverse, since we have $\cD_1(\R)=\{ 0 \}$.
}\end{ex}

\smallskip

\begin{defn}\label{transLatmix}
For $r,s\in \N$ with $r+2s=N$, let $\cD_{r,s}$ denote the locus in the affine space 
$\bA_{r,s}:=\bA^r(\R)\times \bA^s(\C)=\R^r \times \C^s$ given by
\begin{equation}\label{coordD2}
\cD_N=\bigcup_{i=1}^r H_i \cup \bigcup_{j=1}^s L_j\, , \ \ \ \text{ with } \ \  \begin{array}{l} 
H_i =\{ (x_1,\ldots, x_r, z_1, \ldots, z_s)\in \bA^N \,|\, x_i =0 \}\, , \\
L_j=\{ (x_1,\ldots, x_r, z_1, \ldots, z_s)\in \bA^N \,|\, z_j =0 \} \, . \end{array} 
\end{equation}
We say that a lattice $\Lambda \subset \bA_{r,s}\cong \R^N$ is $(r,s)$-mixed transversal if $\Lambda \cap \cD_{r,s}=0$. 
\end{defn}

\smallskip

Let $\bG_m^N =\bA^N \smallsetminus \cD_N$ be the maximal torus in the
affine space $\bA^N$, the complement of the divisor of coordinate hyperplanes,
with $\bG_m$ the multiplicative group, $\bG_m=\GL_1$.
The Haar measure on the torus $\bG_m^N$ is the product of the Haar
measures $dx_i/x_i$ on $\bG_m$, with the volume form
\begin{equation}\label{voltoric}
 \omega_{\bG_m^N}(x)= \frac{dx_1 \wedge \cdots \wedge dx_N}{x_1 \cdots x_N} \, . 
\end{equation} 
The associated volume element is
\begin{equation}\label{volel}
 \upsilon_{\bG_m^N}(x) = \frac{1}{| x_1 \cdots x_N |} \, .
\end{equation}   

\smallskip

In the case where instead of $\R^N$ we consider $\R^r \times \C^s$, we take as volume element
\begin{equation}\label{mixvolel}
 \upsilon_{\bG_m^{r,s}}(x,z): = \upsilon_{\bG_m^r(\R)}(x)\wedge \upsilon_{\bG_m^s(\C)}(z) \wedge \upsilon_{\bG_m^s(\C)}(\bar z)
 = \frac{1}{|x_1| \cdots |x_r| \, |z_1|^2 \cdots |z_s|^2}\, .
\end{equation} 

Transversality of a lattice $\Lambda \subset \R^N$, as in Definition~\ref{transLat}, is equivalently stated as
$\Lambda \smallsetminus \{ 0 \} \subset \bG_m^N$. Thus, the volume form
of $\bG_m^N$ is well defined at all non-zero elements of the lattice $\Lambda$.
The $(r,s)$-mixed transversality condition for $\Lambda \subset \bA^{r,s}$ is the requirement that
the volume element $\upsilon_{\bG_m^{r,s}}$ is finite at all points of $\Lambda \smallsetminus \{ 0 \}$.

\smallskip

\begin{defn}\label{toricpercLat}
The long-range {\em toric percolation} $\cP_\bT(\Lambda)$ on a transverse lattice $\Lambda \subset \R^N$ is
the random graph with vertex set the lattice $\Lambda$ and with edges between
pairs $\{ x,y \}$  included independently at random, 
with inclusion probability 
\begin{equation}\label{Ptoric}
 P_{\beta,\alpha,\bT}(\{x,y\})=1-e^{-\beta J_{\alpha, \bG_m^N}(x-y)} \, , 
\end{equation} 
where the function
$J_{\alpha, \bG_m^N}(x)$, for $x\in \Lambda$, is defined as
\begin{equation}\label{JGm}
J_{\alpha, \bG_m^N}(x) =  \upsilon_{\bG_m^N}(x)^{1+\alpha/N} \, . 
\end{equation}
Similarly, the long-range {\em toric percolation} on a $(r,s)$-mixed transverse 
lattice $\Lambda \subset \bA^{r,s}$ is the random graph with vertex set the lattice 
$\Lambda$ and with edges between
pairs $\{ (x,z), (y,w) \}$  included independently at random, 
with inclusion probability 
\begin{equation}\label{mixPtoric}
 P_{\beta,\alpha,\bT(r,s)}(\{ (x,z),(y,w) \})=1-e^{-\beta J_{\alpha, \bG_m^{r,s}}(x-y,z-w)} \, , 
\end{equation} 
where the function
$J_{\alpha, \bG_m^{r,s}}(x,z)$, for $(x,z)\in \Lambda$, is defined as
\begin{equation}\label{mixJGm}
J_{\alpha, \bG_m^{r,s}}(x,z) =  \upsilon_{\bG_m^{r,s}}(x,z)^{1+\alpha/N} \, . 
\end{equation}
\end{defn}

\smallskip

\begin{rem}\label{1dtoricperc}{\rm
In the $1$-dimensional case with $\Lambda=\Z\subset \R$, the
long range toric percolation $\cP_\bT(\Z)$ agrees with the usual long range
percolation $\cP(\Z)$ with $J_{\alpha, \bG_m}(x)=J_\alpha(x)=|x|^{-1-\alpha}$.
}\end{rem}

\smallskip

\begin{lem}\label{toricPMperc}
For a transverse lattice $\Lambda \subset \R^N$ or a $(r,s)$-mixed transverse 
lattice $\Lambda \subset \bA^{r,s}$, the long-range toric percolation model $\cP_\bT(\Lambda)$
agrees with the power mean long-range percolation model $\cP_{\cM_0}(\Lambda)$ at $t=0$ with
the uniform distribution $\lambda=1/N$. 
\end{lem}

\proof By \eqref{mixJGm}, \eqref{mixPtoric}, and \eqref{MtRCr2s0} we have
\begin{equation}\label{Jalpha0NM0}
 J_{\alpha,0,1/N}(x-y,z-w) = \cM_0 (1/N,(x-y,z-w))^{-N-\alpha} 
 \end{equation}
$$ =\left( \prod_{i=1}^r |x_i-y_i |^{1/N} \, \prod_{j=1}^s |z_j-w_j|^{2/N} \right)^{-N-\alpha}  = \upsilon_{\bG_m^{r,s}}(x-y,z-w)^{1+\alpha/N} \, . $$
 \endproof
 
 \subsection{Comparing toric percolation and power mean percolation} 
 
 Lemma~\ref{toricPMperc} shows that we can view long-range toric percolation on a lattice
 as a special case of power mean long-range percolation at $t=0$. This then implies that one can
 use the behavior of the power mean when changing the parameter $t$ to compare 
 the toric percolation to power mean percolation at other values of $t$. 
 
A first observation is that the arithmetic-geometric mean inequality implies an
estimate relating the inclusion probabilities of the toric percolation model $\cP_\bT(\Lambda)$
and of the usual percolation model $\cP(\Lambda)$.

\begin{lem}\label{amgmmean}
For a transverse lattice $\Lambda \subset \R^N$ or a $(r,s)$-mixed transverse lattice $\Lambda \subset \bA^{r,s}$,
the inclusion probabilities of $\cP_\bT(\Lambda)$ and $\cP(\Lambda)$ satisfy the inequality
\begin{equation}\label{PPtoric}
 \P_{\beta,\alpha,\bT}((x,z),(y,w)) \geq \P_{\beta,\alpha}((x,z),(y,w)) \ \ \  \forall (x,z)\neq (y,w) \in \Lambda \, .
\end{equation} 
In particular, this implies that the respective critical temperatures $\beta_{c,\bT}$ and $\beta_c$ satisfy
 \begin{equation}\label{betacgeq}
 \beta_c \geq \beta_{c,\bT} \, . 
\end{equation} 
\end{lem}

\proof For  $\Lambda \subset \R^N$, the arithmetic-geometric mean inequality \eqref{geoarmean} 
implies that 
\begin{equation}\label{Jgeoar}
 J_{\alpha, \bG_m^N}(x)=(\prod_{i=1}^N |x_i | )^{-1-\alpha/N} \geq 
(\frac{1}{N}\sum_{i=1}^N |x_i |^2 )^{-(N+\alpha)/2} =N^{(N+\alpha)/2} \, \| x \|^{-N-\alpha}=N^{(N+\alpha)/2} \, J_\alpha(x) \, , 
\end{equation}
so that the inclusion probabilities satisfy
\begin{equation}\label{geoarProb}
 1-e^{-\beta\,  J_{\alpha, \bG_m^N}(x-y)} \geq 1-e^{-\beta \cdot N^{(N+\alpha)/2}\, J_\alpha(x-y)} \geq 1-e^{-\beta \, J_\alpha(x-y)}  \, . 
\end{equation} 
Since the long-range toric percolation on a transverse lattice has a higher probability of including an edge 
than the usual long-range percolation on the same lattice, the critical probability satisfies
$\beta_c  \geq  \beta_{c,\bT}$,
since a positive probability of an infinite cluster in the ordinary case implies that the toric case probability is also positive.
The $(r,s)$-mixed case can be treated similarly. In this case, the arithmetic-geometric mean inequality \eqref{geoarmean}
can be combined with the inequality of H\"older means 
\begin{equation}\label{geoarHoldermean}
 (\prod_{i=1}^N |x_i | )^{1/N} \leq \frac{1}{N}\sum_{i=1}^N |x_i |  \leq ( \frac{1}{N}\sum_{i=1}^N |x_i |^2)^{1/2} 
\end{equation} 
to get 
\begin{equation}\label{Holdermean}
(\prod_{i=1}^r |x_i | \cdot \prod_{j=1}^s |z_j| \cdot \prod_{j=1}^s |\bar z_j|  )^{1/(r+2s)} \leq (\frac{1}{(r+2s)^{1/2}} 
( \sum_i |x_i |^2+ 2\sum_j |z_j|^2 )^{1/2} 
\end{equation}
$$ \leq \frac{\sqrt{2}}{(r+2s)^{1/2}} (\sum_i |x_i |^2+\sum_j |z_j|^2)^{1/2} \, . $$
Thus, we obtain
\begin{equation}\label{mixJgeoar}
 J_{\alpha, \bG_m^{r,s}}(x,z)=(\prod_{i=1}^r |x_i | \prod_{j=1}^s |z_j|^2 )^{-1-\alpha/(r+2s)} \geq 
\frac{\sqrt{2}}{(r+2s)^{1/2}}( \sum_{i=1}^{r+2s} |x_i |^2  +\sum_{j=1}^s |z_j|^2)^{-(r+2s+\alpha)/2} 
\end{equation}
$$ =(\frac{(r+2s)}{2})^{(r+2s+\alpha)/2} \| (x,z) \|^{-(r+2s+\alpha)} =(\frac{(r+2s)}{2})^{(r+2s+\alpha)/2} J_\alpha(x,z) \, , $$
so that the inclusion probabilities satisfy
\begin{equation}\label{mixgeoarProb}
 1-e^{-\beta\,  J_{\alpha, \bG_m^{r,s}}(x-y)} \geq 1-e^{-\beta \cdot (s+r/2)^{(r+2s+\alpha)/2}\, J_\alpha(x-y)} 
 \geq 1-e^{-\beta \, J_\alpha(x-y)}  \, . 
\end{equation} 
\endproof

 As observed in \S \ref{PMSec}, the arithmetic-geometric mean inequality is a special case of
 the monotonicity \eqref{monoMtPM} of the power mean function $\cM_t(\lambda,x)$. Thus,
 we can view the comparison of Lemma~\ref{amgmmean} as a special case of a stronger
 comparison result. 
 
 \begin{prop}\label{monoEst}
 For a transverse lattice $\Lambda \subset \R^N$ or a $(r,s)$-mixed transverse lattice $\Lambda \subset \bA^{r,s}$ and
 for any pair $t,t'\in [-\infty,\infty]$ with $t \leq t'$, the inclusion 
 probabilities of the power mean long-range percolation models $\cP_{\cM_t}(\Lambda)$ 
 and $\cP_{\cM_{t'}}(\Lambda)$ satisfy
 \begin{equation}\label{monoprob}
 \bP_{\beta,\alpha,t,\lambda}((x,z),(y,w)) \geq \bP_{\beta,\alpha,t',\lambda}((x,z),(y,w)) 
 \end{equation}
 hence the respective critical inverse temperatures satisfy
  \begin{equation}\label{betactgeq}
 \beta_{c,t,\lambda} \leq \beta_{c,t',\lambda} \, . 
\end{equation} 
 \end{prop}
 
 The proof is analogous to Lemma~\ref{amgmmean}, using the monotonicity \eqref{monoMtPM} 
 instead of the arithmetic-geometric mean inequality. 
 
 \smallskip
 
 We will return to discuss this family of percolation models on lattices in \S \ref{NumFsec}.
 We first need to discuss the other type of percolation models that we will be comparing
 with these models on lattices, namely the hierarchical lattice models. 

\section{The hierarchical lattice percolation model} \label{hierPercSec}

In the previous section we have extended the usual long-range   percolation model
on a lattice to a continuous family of long-range lattice percolation models, depending on
a parameter $t\in [-\infty,\infty]$, which specialize to the usual model at $t=2$ and also includes 
as a special value $t=0$ what we referred to as toric long-range lattice percolation model.

In this section we start with a different type of long-range percolation model, where instead of
a lattice $\Lambda\subset \R^N$, one starts with a {\em hierarchical lattice}. 
The long-range percolation model on the hierarchical lattice is defined in the following way (see  \cite{Hu1}, \cite{Hu2}).

\begin{defn}\label{hierlatdef}{\rm 
Consider chosen integers $N\geq 1$ and $L\geq 2$, take $\bT^N_L:= (\Z/L\Z)^N$. 
The hierarchical lattice $\cH^N_L$ is the 
countable abelian group 
\begin{equation}\label{HNLhlatt}
\cH^N_L :=\oplus_{i=1}^\infty \bT^D_L =
\{ x=(x_1,x_2,\ldots,x_n,\ldots)\in (\bT^N_L)^\N\,|\,
x_i=0 \text{ for all but finitely many } i\geq 1 \}\, .
\end{equation}
It is endowed with the invariant ultrametric
\begin{equation}\label{metricHNL}
\| x-y \|_{\cH^N_L}:= \left\{ \begin{array}{ll} L^{h(x,y)} & x\neq y \\ 0 & x=y \end{array}\right. 
\end{equation}
with 
\begin{equation}\label{hHNL}
h(x,y)=h_{\cH^N_L}(x,y)= \max\{ i\geq 1 \,|\, x_i\neq y_i \}
\end{equation} }
\end{defn}

\begin{rem}\label{lastdig}{\rm
This distance \eqref{metricHNL} computes the {\em last digit} $M$ at which the two sequences $x,y$ differ
and sets $\| x-y \|_{\cH^d_L}=L^M$. We will return to this observation
in \S \ref{FFsec}.
}\end{rem}

\begin{defn}\label{percHNLdef}{\rm
The long-range percolation model $\cP(\cH^N_L)$ on the hierarchical lattice $\cH^N_L$ is the random graph
where an edge $\{ x, y \}$  between $x\neq y\in \cP(\cH^N_L)$ is added with probability
\begin{equation}\label{ProbHNL}
 \bP_{\beta,\alpha,L}(\{ x, y \}) = 1- e ^{-\beta J_{\alpha,L}(x-y)} 
\end{equation}
with
\begin{equation}\label{JalphaHNL} 
J_{\alpha,L}(x-y)=\| x-y \|_{\cH^N_L}^{-N -\alpha} \, . 
\end{equation}
}\end{defn}

\smallskip

For large $\| x- y \|_{\cH^d_L}$ we have 
$$ \bP_{\beta,\alpha,L}(\{ x, y \})\sim \beta  \| x-y \|_{\cH^d_L}^{-d -\alpha}\, . $$
Also note that the function $J_{\alpha,L}$ of \eqref{JalphaHNL} is {\em integrable}, namely it satisfies
\begin{equation}\label{finsumJ}
\sum_x J_{\alpha,L}(x) \sim \sum_N L^N L^{-N(d+\alpha)} <\infty 
\end{equation}
since $\# \{ x\,|\, h_{\cH^d_L}(x,0) = N\}=L^{N-1} (L-1)\sim L^N$. 

We also recall a property of the hierarchical lattice that is different from the case of ordinary lattices and 
that will play a role in the following (Remark~1.3 of \cite{Hu1}).

\begin{rem}\label{hier1D}{\rm
The long-range percolation model $\cP(\cH^N_L)$ on the hierarchical lattice $\cH^N_L$ in dimension $N$ with
parameter $\alpha$ is equivalent to the one-dimensional $\cP(\cH^1_L)$ hierarchical lattice 
long-range percolation model  with parameter $\alpha/N$, so effectively this model is always one-dimensional.
}\end{rem}

\section{Function fields: local models and hierarchical percolation} \label{FFsec}

We reviewed the two settings we want to compare: the long-range percolation models on lattices
(with the one-parameter family generalization we introduced, given by the power mean model)
and the long-range percolation on the hierarchical lattice. We now build a setting, based on the
theory of global and local fields, where these two types of long-range percolation models can
be directly compared. We start in this section with the case of {\em function fields} and how
one can reformulate the long-range percolation model on the hierarchical lattice in this setting.
In \S \ref{NumFsec} we will discuss the case of number fields and the relation to ordinary lattices.
We will show that both settings have non-archimedean components that provide the intermediate
link between the two types of long-range percolation models via the respective adeles rings
and the adelic formula for the norm. 

\smallskip

For a general background on global and local fields, and other number theoretic notions
used in this paper, we refer the reader to \cite{Neuk}, which gives particular
attention to the analogies between number fields and function fields. 

\smallskip

\subsection{Global and local fields}

There are two types of global fields: number fields in characteristic zero, and function fields in
positive characteristic $p>0$. Number fields are extensions of $\Q$ of finite degree $[\K,\Q]=n$,
while the function fields are similarly finite extensions $\F_q(C)$ of the field of rational functions 
$\F_p(\P^1)$, with $q=p^r$ and $C$ an algebraic curve over $\F_p$ (branched cover of $\P^1$). 

There are, correspondingly, two types of non-archimedean local fields, that arise as completions
of global fields. In characteristic zero these are given by $p$-adic fields (finite extensions of the field $\Q_p$),
while in positive characteristic they are fields of formal Laurent series, $K=\F_q((t))$. 

In both cases, local fields can be obtained by the same procedure. 
Given a discrete valuation ring (DVR) $(R,\fm)$, with maximal ideal $\fm=(\varpi)$ for some
generator $\varpi$, one can define on the field of fractions $K$ a discrete valuation
so that $\fm^i=\{ x \in K \,|\, \nu(x) \geq i \}$.  A non-Archimedean metric $d(x,y)=|x-y|$ is defined 
by setting $|x|=\alpha^{v(x)}$ for $0<\alpha<1$. The completion $\hat K$ of $K$ 
with respect to the topology defined by this metric is 
$\hat K=K\otimes_R \hat R=\hat R[1/\varpi]$, where $\hat R=\varprojlim_i R/\fm^i R$.

For example, for $K=\Q$ and $R=\Z_{(p\Z)}$, the localization of $\Z$ at the
prime ideal $p\Z$, we obtain $\hat K=\Q_p$ the field of $p$-adic numbers,
and $\hat R=\Z_p$ the ring of $p$-adic integers, 
with the $p$-adic metric defined as above with $\alpha=1/p$. 
Note that $\Z$ is a lattice in $\R$ (the Archimedean completion of $\Q$)
but is not a lattice in the non-Archimedean completions $\Q_p$, as elements
of $\Z$ accumulate at $0$ in the topology induced by the $p$-adic norm. 

For both types of local field $K$, we will denote by $\cO_K=\{ x\in K \,|\, |x|\leq 1 \}$
the ring of integers, by $\cO_K^\times =\{ x\in K \,|\, |x| =  1 \}$ the units, and by $k=\cO_K/\fm_K$
the residue field (a finite field) with $\fm_K=\{ x\in K \,|\, |x|<1 \}$ the maximal ideas, where 
$\fm_K=(\varpi)$ has a single generator (called a uniformizer). 
In $K^*$ one has a  decomposition $x=\varpi^n u$, with $u$ a unit and $n=\nu(x)$ the 
valuation with $|x|=q^{-\nu(x)}$ for $q=\# \cO_K/\fm_K$. 

Here, we focus on the positive characteristic case of Laurent series and we will return to 
the $p$-adic cases in \S \ref{NumFsec}.

For $K=\F_q((t))$, the field formal Laurent series (expansion at $0$), 
the valuation is given by $\nu_0(f)=M$, the 
smallest integer with nonzero coefficient, namely such that
$$ f(t)=t^M (a_0 + a_1 t + a_2 t^2 + \cdots) $$
with $a_0\neq 0$, and we have $| f |_0=q^{-\nu_0(f)}$. The ring of integers is then  
$\cO_K=\F_q[[T]]$, with $T=1/t$, and the maximal ideal is $\fm_K=(T)$, with  $\cO_K/\fm_K=\F_q$,
the constant functions.  

For this case of positive characteristic, to see the local fields associated to a given
function fields, consider 
a smooth projective curve $C$ over a finite field $\F_q$ of characteristic $p$, with field of functions $\F_q(C)$.
This global field $\F_q(C)$ is a finite extension of the field of rational functions $\F_p(\P^1)=\F_p(t)$:
the realization of $C$ as a branched cover $C\to \P^1$ gives a ramified extension, and extensions
$\F_{p^r}$ of the field of coefficients give unramified extensions. Consider points $x\in C$, with degree $d_x=[k(x):\F_q]$. 
Points of $C$ determine valuations $\nu_x$ (places of the global field), by taking the expansion at $x$ of functions 
$f\in \F_q(C)$. 

In the case of $\P^1$, for $x\in \F_q$, the local field is obtained by considering the 
local Laurent series expansion $\F_q((t-x))$, with  ring of integers 
the power series expansions $\F_q[[ t-x ] ]$. The remaining point of $\P^1$ is the 
point at infinity: for $x=\infty$ we have local field $\F_q((t^{-1}))$. The valuation 
for the point at infinity is $\nu_\infty(f)=-\deg(P(f))$ for the polynomial part $P(f)$ in $\F_q[t]$
of the series $f\in \F_q((t^{-1}))$.
For more general curves $C$, one similarly considers an affine curve $C^{\rm aff}$ and
points at infinity $C\smallsetminus C^{\rm aff}$.

\subsection{Adeles for function fields}\label{FFadelesSec}

For a function field $\K=\F_q(C)$ and a chosen point $\infty\in C\smallsetminus C^{\rm aff}$ 
of degree $d_\infty$, we have
\begin{equation}\label{valinfty}
 | f |_\infty = q^{- d_\infty \nu_\infty(f)} \, . 
\end{equation} 
Let $K=\K_\infty$ be the completion of $\K$ at this point $\infty$ in the metric defined by $|\cdot|_\infty$,
and let $A$ be the ring of functions in $\K$ regular outside of $\infty$. Let  
$\cV_\K$ be the set of places of $\K$ (closed points of $C$) and $\cV_A=\cV_\K\smallsetminus \{ \infty \}$,
with completion $\K_x$ in the $|\cdot|_x$ norm determined by 
$\nu_x(f)={\rm ord}_x(f)$, and with $A_x=\{ f\in \K_x\,|\, |f|_x \leq 1 \}$. 
The ring of adeles is the restricted product $\bA_\K=\prod'_{x\in \cV_\K} \K_x$ where all but finitely many of the 
coordinates $f_x$ of $f=(f_x)\in \bA_\K$ are in $A_x$. The ring of finite adeles is $\bA_{\K,{\rm fin}}=\prod'_{x\in \cV_A} \K_x$,
with maximal compact subring $\hat\cO_\K =\prod_{x\in \cV_A} A_x$.

The norms $|\cdot|_x$ for $x\in \cV_\K$ satisfy the {\em adelic relation}: 
\begin{equation}\label{adelicnormFF}
\prod_{x\in \cV_\K} | f |_x =1 \, , \ \ \ \ \  \forall f\in \K=\F_q(C)\, , \ f\neq 0 \, . 
\end{equation}

This adelic relation for function fields, and the corresponding one for number fields,
will be the key property that establishes the relation between the different percolation
models we are discussing. 

\subsection{Hierarchical lattice and function fields} \label{HlattFFsec}

For a prime $p$ and $q=p^r$, consider the global field of rational functions
$\F_q(\P^1)=\F_q(t)$ and the polynomial subring $\F_q[t] \subset \F_q(t)$. At
the point at infinity $\infty \in \P^1$, we consider the local field $\F_q((t^{-1}))$,
corresponding to the valuation $\nu_\infty(f)=-\deg(P(f))$, with
$P(f)$ the projection onto the polar part, namely the abelian 
subgroup $\F_q[t] \subset \F_q((t^{-1}))$.

\begin{prop}\label{abgrpHNLFqt}
As an {\em abelian group}, $\F_q[t]$ is isomorphic to the hierarchical lattice 
$$ \cH^1_q = \{ a=(a_0,a_1,\ldots, a_k) \,|\, k\geq 0\, , a_i\in \F_q \} =\oplus_{i=0}^\infty \F_q \, . $$
The inclusion probabilities of the long-range percolation model on the hierarchical lattice are then of the form
\begin{equation}\label{ProbPFq}
 \bP_{\beta,\alpha, q,\infty}( \{ f, g \}) = 1- e ^{-\beta J_{\alpha,q,\infty}(f-g)} 
\end{equation}
with
\begin{equation}\label{JPFq}
J_{\alpha,q,\infty}(f-g)=| f-g |_\infty^{-(1+\alpha)}=q^{(1+\alpha)\nu_\infty(f-g)}  \, . 
\end{equation}
\end{prop}

\proof The key identity relating \eqref{ProbPFq} to \eqref{ProbHNL} is
\begin{equation}\label{nuinftyhfg}
h_{\cH^1_q}(f,g)=\max_i \{ a_i(P(f-g))\neq 0 \} = \deg (P(f-g)) =
-\nu_\infty(f-g)  \, ,
\end{equation}
with $h_{\cH^1_q}(f,g)$ defined as in \eqref{hHNL}. We then see that
\begin{equation}\label{JinftyHNL}
J_{\alpha,q}(f-g)=\| f-g \|_{\cH^1_q}^{-(1 +\alpha)} =q^{-(1+\alpha) h_{\cH^1_q}(f,g)} = q^{-(1+\alpha)\deg(P(f-g))}
\end{equation} 
$$  = q^{(1+\alpha) \nu_\infty(f-g)} = | f-g |_\infty^{-(1+\alpha)} =J_{\alpha,q,\infty}(f-g) \, ,  $$
which then gives \eqref{ProbPFq}. 
\endproof

\smallskip

With this reinterpretation, we can view the usual long-range percolation model on the hierarchical lattice
as the component at infinity of a long-range percolation model associated to a function field. We will then
equivalently write $\cP(\cH_q^1)$ or $\cP_\infty(\cH_q^1)$ to denote this percolation model, where
the second notation suggests that, according to Proposition~\ref{abgrpHNLFqt}, we interpret this model
as the component at the infinite place $\infty$ of a percolation model associated to the function field.

\subsection{Function fields and hierarchical lattices: local models at finite places} \label{FFlocalPercSec}

Proposition~\ref{abgrpHNLFqt} shows that the long-range percolation model on the hierarchical lattice
can be described in terms of the local field $\F_q((t^{-1}))$ with the valuation 
$\nu_\infty(f)=-\deg(P(f))$, at the point at infinity $\infty \in \P^1$.  In order to extend this to
a percolation model associated to the global field $\F_q(\P^1)$, or more generally $\F_q(C)$,
we need to consider what such a model looks like at the finite places of the global field, 
namely the closed points $x\in C$, $x\neq \infty$.  

\begin{defn}\label{localFFmod}
Let $\K=\F_q(C)$ be a function field, $\cV_\K$ the set of places of $\K$, with a choice of
a place at infinity $\infty$ and $\cV_A$ the remaining finite places. 
the local long-range percolation model $\cP_x(\cH^1_q)$ at $x\in \cV_A$ is the
random graph with vertices $f \in \F_q[t] \simeq \cH^1_q$ and with inclusion
probabilities
\begin{equation}\label{ProbxH1q}
 \bP_{\beta,\alpha, q,x}( \{ f, g \}) = 1- e ^{-\beta J_{\alpha,q,x}(f-g)} \, ,
\end{equation}
where now we have
\begin{equation}\label{JxH1q}
J_{\alpha,q,x}(f-g)=| f-g |_x^{(1+\alpha)}=q^{-(1+\alpha)\nu_x(f-g)} \, . 
\end{equation}
\end{defn}

First observe the function $J_{\alpha,q,x}(f)$ defined as in \eqref{JxH1q} has very
different properties from the analogous function at the infinite place of \eqref{JPFq}.

\begin{lem}\label{JqxInfSum}
 The function $J_{\alpha,q,x}$ of \eqref{JxH1q} is non-integrable:
$\sum_f J_{\alpha,q,x}(f)=\infty$.
\end{lem}

\proof
The function $J_{\alpha,q,x}(f-g)$ computes the {\em position of the first coefficient} at which $f$ and $g$
differ at $x$, namely the position ${\rm ord}_x(f)$ of the first nonzero coefficient of the expansion at $x$ of $f-g$. 
This implies that the function $J_{\alpha,q,x}(f)$ is non-integrable. Indeed, 
fixing the order ${\rm ord}_x(f)$ at a given closed point $x\in C$ still leaves 
arbitrary possible order at other closed points of $C$, so that $\nu_x(f)=M_x$ 
has infinite multiplicity.
\endproof

Compare Lemma~\ref{JqxInfSum} with Remark~\ref{lastdig} and the integrability of $J_{\alpha,q,\infty}$. 

\begin{prop}\label{infclusterx}
In each local model $\cP_x(\cH^1_q)$ at the finite places $x\in \cV_A$ of the
function field $\K$, one always has an infinite cluster, because the critical
inverse temperature is $\beta_{c,q,x}=0$.
\end{prop}

\proof One can directly see an infinite cluster in the following way. 
As in Proposition~\ref{abgrpHNLFqt}, we identify the hierarchial lattice $\cH^1_q$, as an
abelian group, with
\begin{equation}\label{H1qFqt}
\cH^1_q \cong \F_q[t]  \, .
\end{equation}
For simplicity we focus on the case $C=\P^1$ 
and $x=0$. The other places are treated analogously up to a change of coordinates. 
We write polynomials $f\in \F_q[t]$ in the form $f(f)=B t^M P(t)$, where $M=\nu_0(f)$
and $P(t)$ is a monic polynomial with all zeros at points $x\neq 0$, and with $B \in \F_q$.
We restrict to the subgraph of the random graph $\cP_x(\cH^1_q)$ with
vertices the $f_{B,M}(f):=B t^M P(t)$ with a fixed $P(t)$. The set of vertices of this subgraph
can then be identified with the set of pairs $(B,M)\in \F_q\times \Z^+$ and
the inclusion probability \eqref{ProbxH1q} on this subgraph can be written as 
$$ \P_{\beta,\alpha,q,0} (\{ f_{B,M}, f_{B',M'} \}) =1- e^{-\beta | f_{B,M}-f_{B',M'} |_{x=0}^{(1+\alpha)}}\, . $$
When either $M\neq M'$ or $m=M'$ and $B-B'\neq 0$ in $\F_q$, this is 
$$ \P_{\beta,\alpha,q,0} (\{ f_{B,M}, f_{B',M'} \})= 1- e^{-\beta q^{-(1+\alpha)\min\{ M, M'\}} } \, . $$
Fix $f_{B,M}$ and consider the set $\Omega_N=\{ f_{B',M'} \, |\, M'\geq N \}$, for some $N>M$. The probability of having
no edges between $f_{B,M}$ and $\Omega_N$ is then equal to
$$ \prod_{M'\geq N} e^{-\beta q^{-(1+\alpha)\min\{ M, M'\}} }= e^{-\beta \sum_{M'\geq N} q^{-(1+\alpha)M }} =0 $$
so, with probability one, there is an edge, $\{ f_{B,M}, f_{B',M'} \}$ for some $f_{B',M'}$ in $\Omega_N$. Since we can keep
repeating the same argument with a new set $\Omega_{N'}$ with $N'> M'$, we obtain an infinite cluster.
\endproof 

The adelic product formula \eqref{adelicnormFF} gives a relation between the functions $J_{\alpha,q,x}$ and 
the function $J_{\alpha,q}$ of the hierarchical lattice model.

\begin{lem}\label{lemJxJinfty}
For all $f\in \F_q[t]\simeq \cH^1_q$ with $f\neq 0$, we have
\begin{equation}\label{JxJinf}
\prod_{x\in \cV_A} J_{\alpha,q,x}(f) = J_{\alpha,q}(f) \, ,
\end{equation}
where $\cV_A$ is the set of finite places of the function field $\K=\F_q(C)$.
\end{lem}

\proof The adelic product formula \eqref{adelicnormFF} gives
$$ \prod_{x\in \cV_A} J_{\alpha,q,x}(f) = \prod_{x\in \cV_A} |f|_x^{1+\alpha} = |f|_\infty^{-(1+\alpha)} = J_{\alpha,q,\infty}(f) \, , $$
where the latter is equal to $J_{\alpha,q}(f)$ by Proposition~\ref{abgrpHNLFqt}.
\endproof

\subsection{Function fields and hierarchical lattices: adelic model} 

With the local modesl at the finite places $x\neq \infty$ as in Definition~\ref{localFFmod} 
and the long-range percolation model on the hierarchical lattice at the infinite place $x=\infty$, 
as in Proposition~\ref{abgrpHNLFqt}, we obtain the following adelic model. 

\begin{defn}\label{adelicFFperc}
The adelic long-range percolation model $\cP_{\A_{\K,{\rm fin}}}(\cH^1_q)$ over a 
function field $\K=\F_q(C)$ is the random graph with set of vertices the hierarchical
lattice $\cH^1_q$ (identified as an abelian group with $\F_q[t]$), with inclusion
probabilities
\begin{equation}\label{ProbFFadele}
\P_{\underline{\beta},\underline{\alpha},q, \A_{\K,{\rm fin}}} (\{ f, g \}) =\prod_{x\in \cV_A}  ( 1- e^{-\beta_x J_{\alpha_x,q,x}(f-g)} )
=\prod_{x\in \cV_A}  ( 1- e^{-\beta_x | f-g |_x^{(1+\alpha_x)}} ) \, ,
\end{equation}
where $\underline{\beta}=(\beta_x)_{x\in \cV_A}$ and $\underline{\alpha}=(\alpha_x)_{x\in \cV_A}$.
\end{defn}

We can view this model as the hierarchical lattice $\cH^1_q$ with independent probabilities
$1- e^{-\beta_x J_{\alpha_x,q,x}(f-g)}$ of having an edge connecting $f$ and $g$ in the
$x$-local model. Equivalently we can see this as a random graph with vertex set $\F_q[t]$,
where one randomly adds ``adelic edges"
between vertices $f,g\in \F_q[t]$, which means a collection of edges, included independently
at random, in each of the local models at each finite place $x$. The existence of an edge in 
$\cP_{\A_{\K,{\rm fin}}}(\cH^1_q)$ means the existence of an edge in each $\cP_x(\cH^1_q)$,
for every finite place $x\in \cV_A$.

The inverse parameters $\underline{\beta}=(\beta_x)_{x\in \cV_A}$ and $\underline{\alpha}=(\alpha_x)_{x\in \cV_A}$
are in principle
independently associated to each of the local models at the closed points $x\in C$.
In general, we can make an arbitrary choice of $\underline{\beta}=(\beta_x)_{x\in \cV_A}$,
with the only condition that the infinite product
$$ \prod_{x\in \cV_A} (1-e^{-\beta_x}) $$
converges. This condition is dictated by the fact that, for any choice of $f,g$, 
we have $J_{\alpha_x,q,x}(f-g)=1$ for all but finitely many $x\in \cV_A$.

There is, however, a natural choice that coordinates the parameters $\beta_x$ and $\alpha_x$
and is dictated directly by the geometry of the algebraic curve $C$.

Recall that the zeta function $Z(C,t)$ of a curve $C$ over $\F_q$ can be written in the form
of an Euler product as
\begin{equation}\label{zetaZCs}
Z(C,s)= \prod_x (1- q^{- s \deg(x)})^{-1} \, ,
\end{equation}
with the product over the closed points of $C$ (the places of $\K=\F_q(C)$).

Given a finite subset $S$ of closed points of $C$, we also write
\begin{equation}\label{zetaZCsS}
Z^{(S)}(C,s)= \prod_{x \in \cV_A\smallsetminus S}  (1- q^{- s \deg(x)})^{-1} \, ,
\end{equation}
with the product over the finite places not in $S$.

\begin{defn}\label{betaxZetaDef}{\rm
For a function field $\K=\F_q(C)$ of a curve $C$ over $\F_q$, we take
$\underline{\beta}=(\beta_x)_{x\in \cV_A}$ as
\begin{equation}\label{betaxZeta}
\beta_x = \beta \, \deg(x) \, , \ \ \  \forall x\in \cV_A \, ,
\end{equation}
and we take $\underline{\alpha}=(\alpha_x)_{x\in \cV_A}$ of the form
\begin{equation}\label{alphaxZeta}
\alpha_x = \alpha + \log_q \deg(x) \, , \ \ \  \forall x\in \cV_A  \, . 
\end{equation}
for some fixed $\alpha$ and $\beta$. 
}\end{defn} 

\begin{lem}\label{betaZetaProd}
\begin{enumerate}
\item For $\underline{\beta}=(\beta_x)_{x\in \cV_A}$ and  $\underline{\alpha}=(\alpha_x)_{x\in \cV_A}$ as in
Definition~\ref{betaxZetaDef}, and 
for $f,g\in \F_q[t]\simeq \cH^1_q$, such that $J_{\alpha,q,x}(f-g)$ is small at all finite places
$x\in  \cV_A$ with $\nu_x(f-g)\neq 0$, the inclusion probability \eqref{ProbFFadele} of the
adelic long-range percolation model satisfies
\begin{equation}\label{adProbFFsmallJ}
\P_{\underline{\beta},\underline{\alpha},q, \A_{\K,{\rm fin}}} (\{ f, g \}) \sim Z^{(S_{f-g})}(C,\beta)^{-1} \cdot D_{S_{f-g}} \cdot \beta^{\# S_{f-g}} \cdot 
\prod_{x\in S_{f-g}} | f-g |_x^{(1+\alpha)} \, ,
\end{equation}
for a finite set $S_{f-g}$, with 
\begin{equation}\label{degsCx}
D_{S_{f-g}}:=\prod_{x\in S_{f-g}} \deg(x)^{1-\nu_x(f-g)} \, . 
\end{equation} 
\item For $\underline{\beta}=(\beta_x)_{x\in \cV_A}$ as in Definition~\ref{betaxZetaDef} and constant $\alpha_x=\alpha$ for
all $x\in \cV_A$, we have
\begin{equation}\label{adProbFFsmallJ2}
\P_{\underline{\beta},\alpha,q, \A_{\K,{\rm fin}}} (\{ f, g \}) \sim Z^{(S_{f-g})}(C,\beta)^{-1} \cdot \tilde D_{S_{f-g}} \cdot \beta^{\# S_{f-g}} \cdot 
\prod_{x\in S_{f-g}} | f-g |_x^{(1+\alpha)} \, ,
\end{equation}
where 
\begin{equation}\label{constDStilde}
\tilde D_{S_{f-g}} := \prod_{x\in S_{f-g}} \deg(x) \, .
\end{equation}
\end{enumerate}
\end{lem}

\proof For a given $f\in \F_q[t]$, for all but finitely many $x\in \cV_A$ we have $\nu_x(f)=0$, hence
$J_{\alpha,q,x}(f)=1$. For a given $f$ we denote by $S_f\subset \cV_A$ the finite set of places
where $\nu_x(f)\neq 0$.
Thus, when we compute  $\P_{\underline{\beta},\alpha,q, \A_{\K,{\rm fin}}} (\{ f, g \})$ as in \eqref{ProbFFadele},
we have
$$ \P_{\underline{\beta},\underline{\alpha},q, \A_{\K,{\rm fin}}} (\{ f, g \}) = \prod_{x\in \cV_A\smallsetminus S_{f-g}}  ( 1- q^{-\beta \deg(x)} ) \, \cdot \, \prod_{x\in S_{f-g}}  ( 1- e^{-\beta_x | f-g |_x^{(1+\alpha_x)}} ) $$
$$ = Z^{(S_{f-g})}(C,\beta)^{-1} \cdot \prod_{x\in S_{f-g}}  ( 1- e^{-\beta_x | f-g |_x^{(1+\alpha_x)}} ) \, . $$
If $|f - g|_x$ (hence $J_{\alpha_x,q,x}(f-g)$) is small for $x\in  S_{f-g}$, we then obtain 
$$ \prod_{x\in S_{f-g}}  ( 1- e^{-\beta_x | f-g |_x^{(1+\alpha_x)}} ) \sim  
\beta^{\# S_{f-g}} \cdot  \prod_{x\in S_{f-g}} \deg(x) q^{-\nu_x(f-g) \, \log_q(\deg(x)) } \cdot 
\prod_{x\in S_{f-g}} | f-g |_x^{(1+\alpha)} $$ $$ =
D_{S_{f-g}} \cdot \beta^{\# S_{f-g}} \cdot 
\prod_{x\in S_{f-g}} | f-g |_x^{(1+\alpha)} \, , $$
with $D_{S_{f-g}}$ as in \eqref{degsCx}, hence we get \eqref{adProbFFsmallJ}. 
The second case, with $\alpha_x=\alpha$ constant, is analogous. We have 
$$ \prod_{x\in S_{f-g}}  ( 1- e^{-\beta_x | f-g |_x^{(1+\alpha)}} ) \sim  
\beta^{\# S_{f-g}} \cdot  \prod_{x\in S_{f-g}} \deg(x) \cdot 
\prod_{x\in S_{f-g}} | f-g |_x^{(1+\alpha)}  \, . $$
\endproof

\begin{thm}\label{localadelicFF}
Let $S$ be a finite (large) subset $S\subset \cV_A$ of finite places of a function field $\K=\F_q(C)$.
Let $\underline{\beta}=(\beta_x)_{x\in \cV_A}$ with $\beta_x=\beta \deg(x)$ 
as in \eqref{betaxZeta}. 
For $f,g\in \F_q[t]$ with large $\| f-g \|_{\cH^1_q}$ and such that $S_{f-g}=\{ x\in \cV_A\,|\, \nu_x(f-g)\neq 0 \} \subset S$,
the probabilities of the adelic long-range percolation model $\cP_{\A_{\K,{\rm fin}}}(\cH^1_q)$ behave
like those of the hierarchical model, in the following sense. There are $\beta_{\A,S}, \beta'_{\A,S}$ of the form 
\begin{equation}\label{betaAS}
\beta_{\A,S}= \left\{ \begin{array}{ll}
\frac{\beta^{\# S}}{Z(C,\beta)} & \beta \leq 1 \\[3mm]
\frac{\beta}{Z(C,\beta)} & \beta > 1 
\end{array}\right. \ \ \ \text{ and } \ \ \ 
\beta'_{\A,S}= \left\{
\begin{array}{ll}
\frac{\beta}{Z^{(S)}(C,\beta)} & \beta \leq 1 \\[3mm]
\frac{\beta^{\# S}}{Z^{(S)}(C,\beta)} & \beta > 1
\end{array}\right.
\end{equation}
such that:
\begin{enumerate}
\item for $\underline{\alpha}=(\alpha_x)$ with constant $\alpha_x=\alpha$, 
\begin{equation}\label{ProbAfinH1q}
\P_{\beta_{\A,S},\alpha,q}  (\{ f, g \}) \leq \P_{\underline{\beta},\alpha,q, \A_{\K,{\rm fin}}} (\{ f, g \}) \, ,
\end{equation}
\item for $\underline{\alpha}=(\alpha_x)$ as in \eqref{alphaxZeta},
\begin{equation}\label{ProbAfinH1q2}
\P_{\underline{\beta},\underline{\alpha},q, \A_{\K,{\rm fin}}} (\{ f, g \}) \leq 
\P_{\beta'_{\A,S},\alpha,q}  (\{ f, g \}) \, . 
\end{equation}
\end{enumerate}
\end{thm}

\proof   
For the given $S\subset \cV_A$, we have
\begin{equation}\label{SfS}
\beta^{\# S_{f-g}} \prod_{x\in S_{f-g}} \deg(x)  \geq \left\{ \begin{array}{ll}  
\beta^{\# S} & \text{for }\, \beta \leq 1 \\
\beta & \text{for }\, \beta \geq 1 \, .
\end{array}\right.
\end{equation}
\begin{equation}\label{SfS2}
\beta^{\# S_{f-g}} \prod_{x\in S_{f-g}} \deg(x)^{1-\nu_x(f-g)}  \leq \left\{ \begin{array}{ll}  
 \beta & \text{for }\, \beta \leq 1 \\
 \beta^{\# S} & \text{for }\, \beta \geq 1 \, .
\end{array}\right.
\end{equation}
Moreover, we have for $\beta>0$
\begin{equation}\label{ZetaSest}
Z(C,\beta)^{-1} \leq Z^{(S_{f-g})}(C,\beta)^{-1} \leq Z^{(S)}(C,\beta)^{-1} \, ,
\end{equation}
since 
$$ Z(C,\beta)^{-1}=\prod_{x\in S_{f-g}} (1-q^{-\deg(x)\beta}) \cdot \prod_{x\in S_{f-g}^c} (1-q^{-\deg(x)\beta}) 
\leq Z^{(S_{f-g})}(C,\beta)^{-1} = \prod_{x\in S_{f-g}^c} (1-q^{-\deg(x)\beta})  \, , $$
while for $S_{f-g}\subset S$ we have
$$ Z^{(S)}(C,\beta)^{-1}=\prod_{x\in S^c} (1-q^{-\deg(x)\beta}) \geq Z^{(S_{f-g})}(C,\beta)^{-1}
=\prod_{x\in S^c_{f-g}} (1-q^{-\deg(x)\beta})\, . $$ 
When $\| f-g \|_{\cH^1_q}$ is
large, $J_{\alpha,q}(f)$ is small, hence for any given $\beta$
$$ \P_{\beta,\alpha,q}  (\{ f, g \}) \sim \beta J_{\alpha,q}(f-g)\, . $$
By Proposition~\ref{abgrpHNLFqt}, this is equal to 
$$  \beta J_{\alpha,q,\infty}(f-g) = \beta \, | f-g |_\infty^{-(1+\alpha)} \, . $$
By Lemma~\ref{lemJxJinfty}, this is then equal to 
$$ \beta \, | f-g |_\infty^{-(1+\alpha)}  = \beta \, \prod_{x\in \cV_A}  | f-g |_x^{(1+\alpha)} = \beta \prod_{x \in S_{f-g}}  | f-g |_x^{(1+\alpha)} = \beta \prod_{x \in S_{f-g}}  J_{\alpha,q,x}(f-g) \, . $$
For $\beta =\beta_{\A,S}$ as in \eqref{betaAS}, and small $J_{\alpha,q,x}(f-g)$ on $S_{f-g}\subset S$, we then have  
$$ \beta_{\A,S} \prod_x J_{\alpha,q,x}(f-g) \leq \frac{\beta^{\# S_{f-g}} \prod_{x\in S_{f-g}} \deg(x)}{Z^{(S_{f-g})}(C,\beta)} \prod_{x\in S_{f-g}}  J_{\alpha,q,x}(f-g) $$ $$ =\prod_{x\in \cV_A\smallsetminus S} (1-e^{-\beta_x J_{\alpha,q,x}(f-g) }) \cdot 
\beta^{\# S_{f-g}} \prod_{x\in S_{f-g}} \deg(x) \,  J_{\alpha,q,x}(f-g) $$
$$ = \prod_{x\in \cV_A\smallsetminus S} (1-e^{-\beta_x J_{\alpha,q,x}(f-g) }) \cdot \prod_{x\in S_{f-g}} \beta_x \,  J_{\alpha,q,x}(f-g) 
\sim \prod_x \bP_{\beta_x,\alpha, q,x}( \{ f, g \}) = \P_{\underline{\beta},\alpha,q, \A_{\K,{\rm fin}}} (\{ f, g \}) \, . $$
Thus, in the first case, with constant $\alpha_x=\alpha$, we obtain
$$ \P_{\beta_{\A,S},\alpha,q}  (\{ f, g \}) \leq \P_{\underline{\beta},\alpha,q, \A_{\K,{\rm fin}}} (\{ f, g \})\, . $$
Similarly, in the second case, with $\underline{\alpha}=(\alpha_x)$ as in \eqref{alphaxZeta}, 
we have, for small $J_{\alpha_x,q,x}(f-g)$ on $x\in S_{f-g}\subset S$,
$$ \P_{\underline{\beta},\underline{\alpha},q, \A_{\K,{\rm fin}}} (\{ f, g \}) = 
\prod_{x\in \cV_A\smallsetminus S_{f-g}} (1-e^{-\beta_x J_{\alpha_x,q,x}(f-g) }) \cdot 
\prod_{x\in S_{f-g}} (1-e^{-\beta_x J_{\alpha_x,q,x}(f-g) }) $$
$$ \sim Z^{(S_{f-g})}(C,\beta)^{-1} \cdot \prod_{x\in S_{f-g}} \beta_x J_{\alpha_x,q,x}(f-g) 
= \frac{\beta^{\# S_{f-g}} \prod_{x\in S_{f-g}} \deg(x)^{1-\nu_x(f-g)}}{Z^{(S_{f-g})}(C,\beta)} \cdot \prod_{x\in S_{f-g}} J_{\alpha,q,x}(f-g) \, . $$
$$  \leq \beta'_{\A,S} \prod_{x\in S_{f-g}} | f-g |_x^{(1+\alpha)} = \beta'_{\A,S} \prod_{x\in \cV_A} | f-g |_x^{(1+\alpha)}
= \beta'_{\A,S}  \, |  f-g |_\infty^{-(1+\alpha)} $$ $$ = \beta'_{\A,S}  J_{\alpha,q}(f-g) \sim \P_{\beta'_{\A,S},\alpha,q}  (\{ f, g \})   \, , $$ 
so that we also obtain
$$ \P_{\underline{\beta},\underline{\alpha},q, \A_{\K,{\rm fin}}} (\{ f, g \}) \leq \P_{\beta'_{\A,S},\alpha,q}  (\{ f, g \})\, . $$
Note that the expressions $\beta_{\A,S}, \beta'_{\A,S}$ of of \eqref{betaAS} are finite for any $\beta>0$, since
the zeta function $Z(C,\beta)$ is a rational function of the variable $T=q^{-\beta}$ with
$$ Z(C,\beta) = \frac{P(C,q^{-\beta})}{(1-q^{\beta}) (1-q^{-\beta +1})} $$
where $P(C,T)$ is a polynomial with integer coefficients, with $P(C,0)=1$ and $P(C,1)$ the number
of line bundles of degree zero on the curve $C$ (see \S 3 of \cite{Mustata}), and 
$$ P(C,T) = \prod_{i=1}^{2g} (1-\omega_i T) $$
with $g=g(C)$ the genus of the curve $C$ and with algebraic integers $\omega_i$ with $|\omega_i|=q^{1/2}$. 
\endproof

We then obtain the following rephrasing of the question on the existence of an infinite cluster
in the hierarchical lattice long-range percolation model.

For a given value of $\alpha >0$, the existence of an infinite cluster at given $\beta$ with $\underline{\beta}=(\beta_x)$
with $\beta_x=\beta \deg(x)$, in the adelic model of Definition~\ref{adelicFFperc} is equivalent to the existence of
a choice of infinite clusters $K_x$ in each local model (as in
Definition~\ref{localFFmod}) at each finite place $x\in \cV_A$, with inverse temperature $\beta_x$, 
with the property that $K=\cap_x K_x$ is still an infinite cluster. 

The existence of infinite clusters $K_x$, for any $\beta_x>0$, is guaranteed by  Proposition~\ref{infclusterx}.
However, for a given collection $\{ K_x \}$ of such infinite clusters, 
whether or not $K=\cap_x K_x$ is also infinite depends on both $C$ and $\beta$.

We can use Theorem~\ref{localadelicFF} to derive conditions on the existence 
of an infinite cluster in the adelic model on the basis of the 
hierarchical lattice of Definition~\ref{percHNLdef}. 

\begin{prop}\label{betaAScritical}
Let $\underline{\beta}=(\beta_x)_{x\in \cV_A}$ with $\beta_x=\beta \deg(x)$ and 
$\underline{\alpha}=(\alpha_x)_{x\in \cV_A}$ with either constant $\alpha_x=\alpha$ or
with $\alpha_x = \alpha + \log_q \deg(x)$ as in \eqref{betaxZeta}. Let $\beta_c$ denote
the critical inverse temperature of the hierarchical lattice $\cH^1_q$, with inclusion
probabilities $\P_{\beta,\alpha,q}(\{ f, g \})$. 
\begin{enumerate}
\item In the case of
constant $\alpha_x=\alpha$, for any $\beta$ satisfying the inequalities
$$ \beta \geq \max\{ 1, \beta_c \, Z(C,\beta) \}\, , $$
there is an infinite cluster with non-zero probability in the adelic model with
inclusion probabilities $\P_{\underline{\beta},\alpha,q, \A_{\K,{\rm fin}}} (\{ f, g \})$. 
\item In the case with $\alpha_x = \alpha + \log_q \deg(x)$, for any $\beta$ satisfying the inequalities
$$ \beta \leq \min\{ 1, \beta_c \, Z(C,\beta) \} $$
there is no infinite cluster with non-zero probability in the adelic model with
inclusion probabilities $\P_{\underline{\beta},\underline{\alpha} ,q, \A_{\K,{\rm fin}}} (\{ f, g \})$.
\end{enumerate}
\end{prop}

\proof The two cases follow directly from the estimates \eqref{ProbAfinH1q} and \eqref{ProbAfinH1q2}
and the conditions
$$ \beta_{\A,S} > \beta_c \ \ \ \text{ and } \ \ \  \beta'_{\A,S} < \beta_c $$
for $\beta_{\A,S}, \beta'_{\A,S}$ as in \eqref{betaAS} and the estimate $Z(C,\beta)^{-1}\leq Z^{(S)}(C,\beta)^{-1}$.
\endproof

\smallskip

We have realized the long-range percolation model on the hierarchical lattice $\cP(\cH^1_q)$
as an adelic model over a function field, where the finite components are the simpler models $\cP_x(\cH^1_q)$.
We can now return to consider the long-range percolation models on a lattice, and their one-parameter
family of deformations introduced in \S \ref{PMtSec} and show how they relate to $\cP(\cH^1_q)$, through
another adelic construction over number fields, where the percolation model at the finite places will
be comparable to the local models $\cP_x(\cH^1_q)$.

\section{Number fields: local models and lattice percolation} \label{NumFsec}

We now return to discuss the percolation models on lattices introduced in \S \ref{PMtSec}, and we
provide a framework where these can be directly compared, through a common underlying
adelic geometry, to the hierarchical lattices discussed in the previous section.

\subsection{Number fields and lattices}

In the following, we will focus our attention on lattices that
arise as rings of integers, or more generally as fractional
ideals inside number fields. We view them as lattices in
a Euclidean space through the set of embeddings given 
by the archimedean places of number fields.

\smallskip

Let $\K$ denote a number field, an algebraic extension of $\Q$ of some degree
$[\K:\Q]=n$. The degree of the extension is the dimension of $\K$ as a
$\Q$-vector space. 
The ring of integers $\cO_\K$ is a free $\Z$-module of rank $n$,
on an integral basis $\{ \omega_1, \ldots, \omega_n \}$. 

Let $\cV_\K$ be the set of places of $\K$, with $\cV_{\K,f}$ the set of finite (non-Archimedean) 
places and $\cV_{\K,\infty}$ the set of infinite, Archimedean, places, namely
the embeddings $\sigma: \K \hookrightarrow \C$. Let $r$ denote the number
of real embeddings and $s$ the number of complex conjugate pairs of  
complex embeddings, so that $n=r + 2 s$. We denote by
$\K_\R:= \R \otimes_\Q \K$. This satisfies
$$ \K_\R \simeq \R^{r} \times \R^{2 s} = \R^n \, , $$
and $\cO_\K$ with its integral basis defines a lattice in $\K_\R$. Similarly,
any fractional ideal $\cI$ in $\cO_\K$ gives a lattice in $\K_\R$ with
${\rm covol}(\cI)= N(\cI) \cdot \sqrt{ {\rm disc}(\cO_\K)}$, with 
$N(\cI)$ the norm and ${\rm disc}(\cO_\K)$ the discriminant (see \S 5 of \cite{Marcus}).

\begin{ex}\label{cycloex}{\rm
Given a primitive $n$-th root of unity $\zeta_n$, the cyclotomic field $\Q(\zeta_n)$
is a number field of degree $[\Q(\zeta_n):\Q]=\varphi(n)$, where $\varphi$
is the Euler totient function, with $\varphi(n)$ the number of primitive $n$-th
roots of unity. The ring of integers in this case is $\cO_{\Q(\zeta_n)}=\Z[\zeta_n]$
with an integral basis given by the primitive roots of unity. This
gives a lattice $\Z[\zeta_n]\simeq \Z^{\varphi(n)} \subset \R^{\varphi(n)}\simeq \Q(\zeta_n)_\R$. 
}\end{ex}

\smallskip
\subsection{Local fields and $p$-adic extensions}\label{pLocFieldsSec}

The p-adic field $\Q_p$ is the completion of $\Q$ in the non-archimedean metric $| x-y |_p$ 
with $|x|_p=p^{-\nu(x)}$, for $x=m/n=p^{\nu(x)} a/b$, with $p\!\mathrel{\not\vert} a,b$. 
The ring of integers $\Z_p=\{ x\in \Q_p\,|\, |x|_p \geq 1 \} \subset \Q_p$ is given by the
projective limit (ordered by divisibility)
\begin{equation}\label{Zpprojlim}
\Z_p =\varprojlim_n \Z/p^n \Z .
\end{equation} 
As an abelian group, $\Z_p$ is Pontrjagin dual to the Pr\"ufer $p$-group 
\begin{equation}\label{Prufer}
Pr(p):= \Z[1/p]/\Z = \varinjlim \Z/p^n\Z \, , 
\end{equation}
where the direct system is ordered by the inclusions. This in turn can be identified with 
the group of all roots of unity $\exp(2\pi i m/p^k)$, for $0\leq m<p^k$, $k\in \Z^+$.

\smallskip

\begin{rem}\label{Znolatt}{\rm
Note that $\Z$ is a lattice in $\R$ (the Archimedean completion of $\Q$)
but is not a lattice in the non-Archimedean completions $\Q_p$, as elements
of $\Z$ accumulate at $0$ in the topology induced by the $p$-adic norm. }
\end{rem}

\smallskip

More generally, let $K$ be a $p$-adic field, that is, a finite extension of $\Q_p$,
with $[K:\Q_p]=n$, 
and let $\cO_K$ be its ring of integers. The discrete valuation $v_K$ associated
to the discrete valuation ring $\cO_K$ determines the $p$-adic absolute value on $K$ as
$$ | x |_p := p^{-v_K(x)/e_K}\, , $$
where $e_K=v_K(p)$ is the ramification index of $K$ over $\Q_p$. 
The ramification index $e_K$ has the property that 
$p \cO_K =\fm_K^{e_K}=(\varpi^{e_K})$. 

\smallskip

The ring of integers is characterized by $\cO_K=\{ x\in K\,|\, |x|_p\leq 1\}$ and
the maximal ideal of $\cO_K$ is $\fm_K=\{ x\in K\,|\, |x|_p < 1\}$. 
The ring of integers $\cO_K$ is a free $\Z_p$-module of rank $n$. 
Let's denote by $\varpi_K$ a generator of the maximal ideal, $\fm_K=(\varpi_K)$. 
When $K=\Q_p$ we have residue field $\cO_K/\fm_K=\F_p$, and
similarly for an extension with $[K:\Q_p]=n$ we have $\cO_K/\fm_K=\F_q$,
with $q=p^f$, 
a finite field extension of $\F_p$ with $f_K:=f=[\F_q : \F_p]$ the inertial degree. 
Then $[K:\Q_p]=n=e_K\cdot f_K$. 

\smallskip

If $K$ and $L$ are two $p$-adic fields with $K \hookrightarrow L$, then
one defines $e_{L/K}$ by the relation $\varpi_K \cO_L =(\varpi_L^{e_{L/K}})$,
and $f_{L/K}=[\cO_L/\fm_L : \cO_K/\fm_K ]$, so that $e_L=e_K\cdot e_{L/K}$
and $f_L=f_K \cdot f_{L/K}$, and $[L:K]=e_{L/K}\cdot f_{L/K}$. This gives
the compatibility of the extensions of the $p$-adic valuation. 

\smallskip

The $p$-adic norm on $p$-adic fields can also be described in
terms of norms of field extensions. Given $K \subset L$ of degree
$[L:K]=n$, viewing $L$ as an $n$-dimensional vector space over $K$,
we have $N_{L/K}(x) = \det(M_x)$ where $M_x: L \to L$ is the $K$-linear map
given by $M_x: y \mapsto xy$ in $L$. Then the $p$-adic norms in $L$ and
$K$ are related by $| x |_p = | N_{L/K}(x) |_p^{1/[L:K]}$.

\subsection{The $p$-adic expansions}\label{padicExpandSec}

For $K=\Q_p$, the p-adic expansion of $x\in \Q_p$ is of the form 
\begin{equation}\label{pExpandQp}
x=p^{-r} (x_0+x_1 p + x_2 p^2 + \cdots x_\ell p^\ell + \cdots )\, ,
\end{equation}
with $x_i \in \{ 0, \ldots, p-1\}=\F_p$. The dense 
field $\Q\subset \Q_p$ of rational numbers is characterized as those points that
have an eventually periodic $p$-adic expansion, and a point $x\in \Q$
has a $p$-adic expansion that is eventually zero iff it is of the form $x=m p^{-\ell}$
for some non-negative integers $\ell,m \in \Z^+$. This identifies the set of 
terminating sequences $x=x_0+x_1 p + x_2 p^2 + \cdots x_\ell p^\ell$ with $\Z^+$
with its monoid structure. Note that  negative integers have infinite $p$-adic expansion, with
$$ -1 = \sum_{j\geq 0} (p-1) p^j \ \ \  \text{ all } x_j=p-1\, . $$
In fact, all negative integers have the $x_j$ eventually equal to $p-1$.
Also note that the addition on $p$-adic expansions is not the coordinatewise addition 
on the $x_i\in \F_p$:  it is the addition of base $p$ expansions, where there is a carry. 

In the more general case of a $p$-adic field $K$ with $[K,\Q_p]=n$, with $\fm_K=(\varpi_K)$, the $p$-adic
expansion \eqref{pExpandQp} for elements in $\Q_p$ is replaced by a unique expansion 
for elements $x\in K$ of the form
\begin{equation}\label{piexpand}
 x= \varpi_K^{-r} (x_0 + x_1 \varpi_K + x_2 \varpi_K^2 + \cdots + x_\ell \varpi_K^\ell + \cdots ) \, , 
\end{equation} 
where the $x_i$ are in a chosen set of representatives of $\cO_K/\fm_K$, identified with
the elements of the finite residue field $\F_q$. We still refer to \eqref{piexpand} as the
$p$-adic (or more appropriately $\varpi_K$-adic) expansion of $x$. 

In this more general setting, characterizing terminating (eventually equal to $0$) and 
eventually periodic expansions is more delicate. It is shown in \cite{ChenHuang} that, for a number field $\K$
of degree $[\K:\Q]=n$, and a principal prime ideal $\wp$ of $\K$, with $\wp=\varpi \cO_\K$, such
that the norm satisfies $N(\wp)> 2^n$, all the elements in the ring of integers $\cO_\K$ of the
number field have a finite or eventually periodic $\wp$-adic expansion, which is a generalization of the
expansion of the form \eqref{piexpand}, in which different choices $\varpi_i$ of the generator
of $\wp$ (differing by multiplication by units of $\cO_\K$) can be used at each term of the
expansion. 

If $\K$ is a number field and $K=\K_\nu$ is the completion at a non-archimedean 
place $\nu\in \cV_{\K,fin}$ associated to a prime $\wp$ of $\K$, then we have
$\cO_\K/\wp \simeq \cO_K/\fm_K =\F_q$, and the generator $(\varpi_K)=\fm_K$
satisfies $\varpi_K\in \wp$.

The $\wp$-adic expansions are then of the form
\begin{equation}\label{wpexpand} 
x= x_0 + x_1 \omega_1 + \cdots + x_\ell \omega_\ell + \cdots
\end{equation}
with $x_i$ in a set of representatives of $\cO_\K/\wp \simeq \cO_K/\fm_K =\F_q$ for $K=\K_\nu$, 
the local field  at the place $\nu\in \cV_\K$ determined by $\wp$, and with
$$ \omega_i =\prod_{j=1}^i \varpi_j \, , \ \ \ \text{ with } \ \  \varpi_j = u_j \, \varpi $$
for some $u_j$ in the group of units of $\cO_\K$, and for $\varpi=\varpi_K$. In the cases where all 
the $u_j=1$ one obtains an expansion of the form \eqref{piexpand}, in the local field 
$K=\K_\nu$.  

\smallskip

Since $N(\wp)=p^{f_\nu}=p^{[\F_q:\F_p]}$ with $\F_q=\cO_\K/\wp$, the condition $N(\wp)>2^{[\K:\Q]}$ corresponds
to primes with 
$$ \frac{\log p}{\log 2} > \frac{ [\K:\Q] }{[\F_q: \F_p]} \, . $$
On the other hand, the principal prime ideals $\wp$ of $\K$ are the prime ideals that split completely in 
the Hilbert class field, the maximal abelian unramified extension of $\K$. For number fields of 
class number $h_\K=1$ the ring of integers $\cO_\K$ is a PID and all prime ideals are principal. 

\smallskip

Thus, in general one cannot assume that the conditions of \cite{ChenHuang} are satisfied, 
to ensure that all elements of $\cO_\K$ have terminating or eventually
periodic $\wp$-expansion, at all non-Archimedean places of $\K$. Therefore, we
proceed in the following way to construct adelic percolation models for number fields.

\begin{defn}\label{defOKplus}
For a number field $\K$ and a prime $\wp$ of $\K$, with $\nu\in \cV_{\K,{\rm fin}}$ the place determined by $\wp$,
we denote by 
$\Sigma_{\K,\wp} \subset \cO_{K_\nu}$ the set of elements with terminating $\wp$-adic expansion,
\begin{equation}\label{OKplus}
\Sigma_{\K,\wp} :=\{ x\in \cO_{K_\nu} \,|\, x= x_0 + x_1 \omega_1 + \cdots + x_\ell \omega_\ell , \, \text{ for some } \ell \geq 0 \} \, .
\end{equation} 
We also denote by $\bar\Sigma_{\K,\wp}\subset \cO_{K_\nu}$ the set of elements with eventually periodic $\wp$-adic expansion.
\end{defn}

\begin{lem}\label{periodic}
For a number field $\K$ and a prime ideal $\wp$ of $\K$,  
elements $x\in \bar\Sigma_{\K,\wp}$ can
be identified with pairs $(a,b)$ in the product $\Sigma_{\K,\wp}\times \Sigma_{\K,\wp}$, where 
$| a |_\nu = | x |_\nu$, if $a\neq 0$, and $| b |_\nu \leq | x |_\nu$, where $\nu \in \cV_{\K,{\rm fin}}$ 
is the place of $\K$ determined by $\wp$.
\end{lem}

\proof We can write an element of $\bar\Sigma_{\K,\wp}$ in the form $x=a\bar{b}$ with
$(a,b)\in \Sigma_{\K,\wp}\times \Sigma_{\K,\wp}$, where $b$ is the period. This means,
in terms of $\wp$-expansions, 
\begin{equation}\label{xperiodic}
 x = a_0 + \sum_{i=1}^\ell a_i u_i \varpi^i + \frac{1}{1-u \varpi^m} \sum_{i=\ell+1}^{\ell+m} b_i u_i \varpi^i 
\end{equation} 
where
\begin{equation}\label{abperiodic}
 a=  a_0 + \sum_{i=1}^\ell a_i u_i \varpi^i  \ \ \ \text{ and } \ \ \ b= \sum_{i=\ell+1}^{\ell+m} b_i u_i \varpi^i \, . 
\end{equation} 
with the $u, u_i$ in the group of units of $\cO_\K$. 
The norm $| x |_\nu$ at the non-Archimedean place determined by $\wp$ is given by
$| x |_\nu =| N_{\K_\nu/\Q_p}(x) |_p =q^{-v_\nu(x)}$ for $\nu |p$, where 
$v_\nu(x)=\max\{ \ell \geq 1 \,|\, x \in \fm^{\ell}_K \}$, for $K=\K_\nu$. Thus, $v_\nu(x)$ is the 
largest $\varpi_K^r$ that divides $x$. So if $a\neq 0$, we have $r\leq \ell$ and $v_\nu(x)=v_\nu(a)$
and $v_\nu(b) \geq v_\nu(x)$ with equality if $a=0$.
\endproof

\begin{defn}\label{defHKwp}{\rm
For a number field $\K$, and $\wp$ any prime ideal of $\K$, let 
\begin{equation}\label{HKwp}
\cH_{\K,\wp}:= \cO_\K \cap \bar\Sigma_{\K,\wp} \subset \cO_{K_\nu} \, ,
\end{equation}
with $\nu \in \cV_{\K,{\rm fin}}$ 
is the place of $\K$ determined by $\wp$. We also define
\begin{equation}\label{HKad}
\cH_\K:=\bigcap_{\wp} \cH_{\K,\wp} \subset \cO_\K \, , 
\end{equation}
as the part of $\cO_\K$ that has finite or eventually periodic $\wp$-expansions at every $\wp$ of $\K$.
}\end{defn}

\begin{rem}\label{QZH}{\rm 
In the case of $\K=\Q$ the sets $\cH_{\Q,p}$ and $\cH_\Q$ are all equal to $\Z$. }
\end{rem}

\subsection{The $p$-adic hierarchical lattice}\label{pHierLattSec}

We now construct a $p$-adic version of the local long-range percolation model at finite places
that we introduced in \S \ref{FFlocalPercSec} in the case of function fields.

\begin{defn}\label{NumFlocmod}
Let $\K$ be a number field such that $\cH_\K$, as in \eqref{HKad}, is a countably infinite set and let $\cV_{\K,{\rm fin}}$
be the set of non-Archimedean places of $\K$. The local long-range percolation model $\cP_\nu(\cH_\K)$,
with $\nu\in  \cV_{\K,{\rm fin}}$ determined by a prime $\wp$ of $\K$, is the random graph with set of vertices 
$\cH_{\K,\wp}$ and with inclusion probabilities 
\begin{equation}\label{ProbNumFlocmod}
\P_{\beta,\alpha,\nu}(\{ x, y \}) = 1- e^{-\beta J_{\alpha,\nu}(x-y)} \, , 
\end{equation}
with
\begin{equation}\label{JNumFlocmod}
J_{\alpha,\nu}(x-y) := | x-y |_\nu^{(1+\alpha)} = q^{-(1+\alpha) v_\nu(x-y)} \, . 
\end{equation}
\end{defn}

This model behaves in the same way as the local models at finite places of a function field
discussed in  \S \ref{FFlocalPercSec}. In particular, the function $J_{\alpha,\nu}$ of \eqref{JNumFlocmod}
does not satisfy the integrability condition, as in Lemma~\ref{JqxInfSum}, and the critical inverse temperature is $\beta_{c,\nu}=0$,
namely there is always an infinite cluster, as in Proposition~\ref{infclusterx}. 
In fact, there is a more geometric way to describe these
local models, in terms of Bruhat--Tits trees. We discuss it in \S \ref{BTtreeSec} for the $p$-adic
case. An analogous formulation with Bruhat--Tits trees can be similarly obtained for
function fields.

\subsection{The local models and the geometry of Bruhat--Tits trees} \label{BTtreeSec} 

\begin{center}
 \begin{figure}[h]
\includegraphics[scale=0.5]{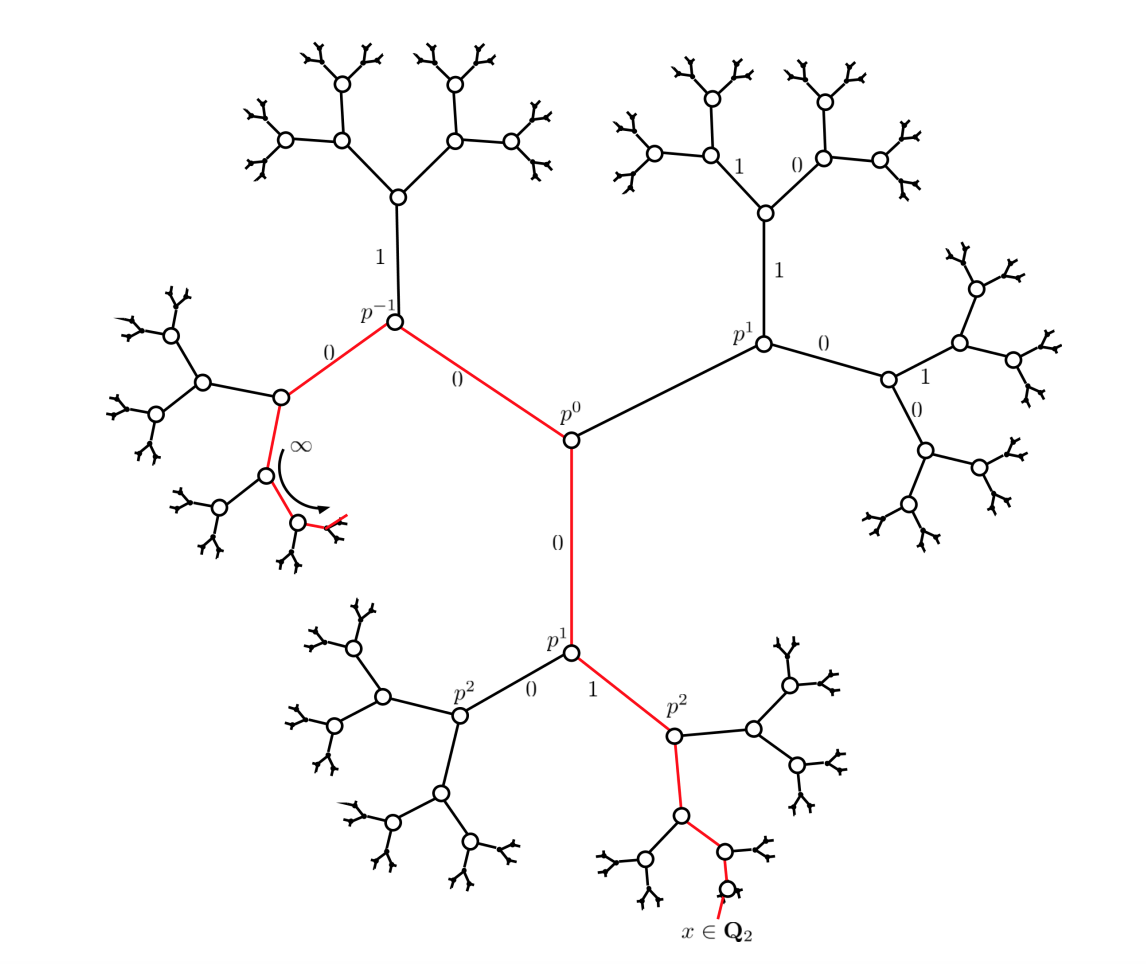}
\caption{The Bruhat--Tits tree of $\Q_2$, from \cite{HMSS}. \label{FigBTtree}}
\end{figure}
\end{center}

One can associate to a $p$-adic field $K$ its Bruhat--Tits tree $\cT_K$. It is the homogeneous
space $\cT_K=\PGL(2,K)/\PGL(2,\cO_K)$, which can be identified with an infinite tree with vertices
of valence $q+1$ with $q=\# \cO_K/\fm_K$. The boundary at infinity of $\cT_K$ is identified with
$\P^1(K)$. The choice of a projective coordinate on $\P^1(K)$ (equivalently, the choice of three
points identified with $\{ 0, 1, \infty \}$) determines a unique vertex in the interior of the tree $\cT_K$,
which is the meeting point of the infinite lines connecting these boundary points pairwise. We
consider that vertex as a root of the tree $\cT_K$. The edges adjacent to the root vertex are labelled by
the elements of $\P^1(\F_q)$ with $\F_q=\cO_K/\fm_K$, where we identify with $\infty \in \P^1(\F_q)$
the edge pointing in the direction of the boundary point $\infty \in \P^1(K)$, see Figure~\ref{FigBTtree}.
We refer the reader to \cite{Man} for a general overview of Bruhat--Tits trees and $p$-adic geometry.

The part of the boundary that is obtained by following paths from the root that starts in any
of the $q$ directions in $\F_q$ gives a copy of $\cO_K$. After going a number of
steps from the root along the path from the root vertex to $0\in \P^1(K)$, the part of the
boundary that can be reached continuing with arbitrary paths away from the root
gives copies of the successive $\fm_K^\ell$, $\ell\geq 1$, as a system of neighborhood of 
$0\in K$. 

 \begin{center}
 \begin{figure}[h]
\includegraphics[scale=0.4]{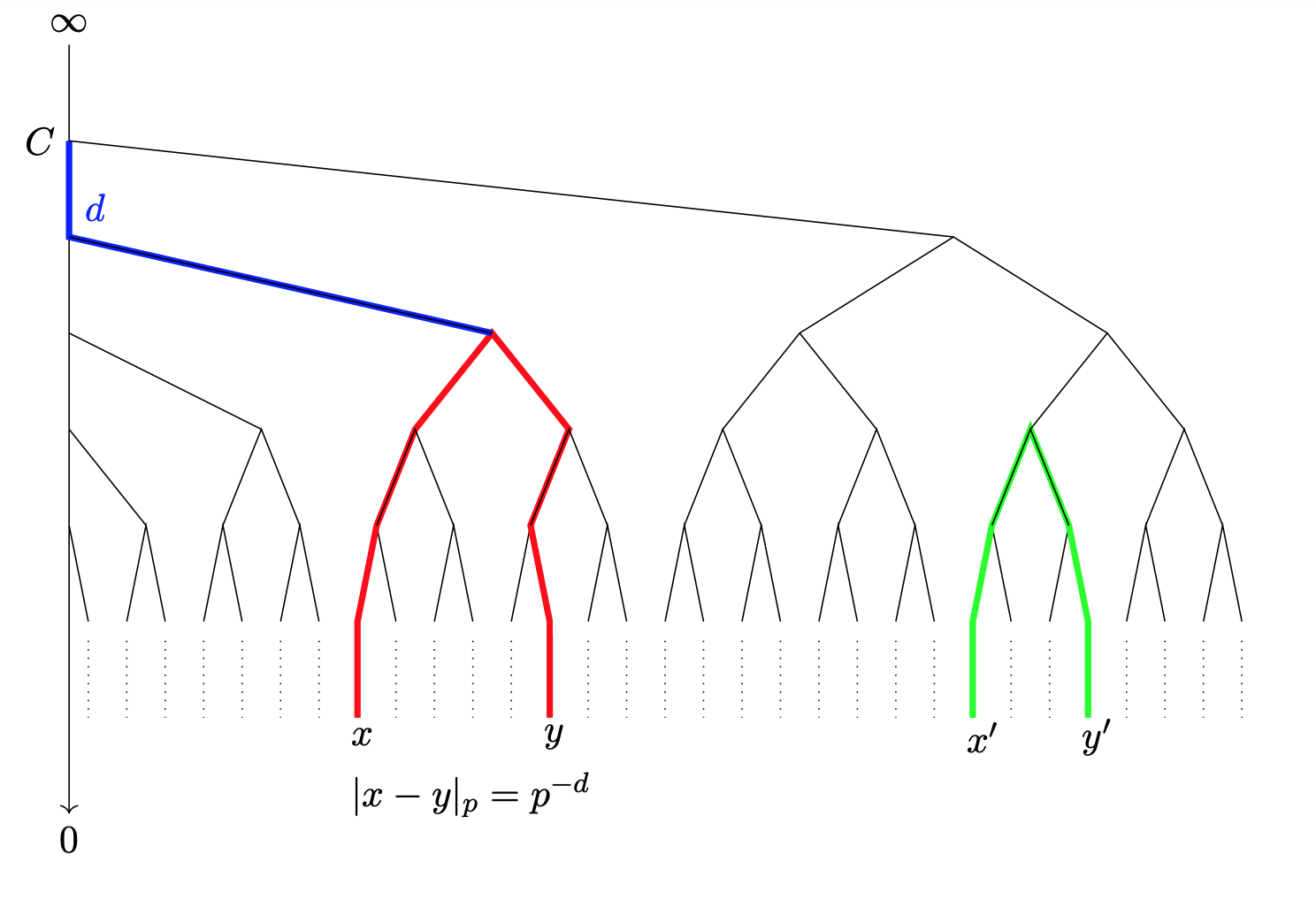}
\caption{The $p$-adic norm and the Bruhat--Tits tree, from \cite{HMSS}. \label{FigBTnorm}}
\end{figure}
\end{center}

The $p$-adic valuation $\nu_K(x-y)$ for $x,y\in K$ is computed by the number $d_\perp(x,y)$ of steps
on the tree that connect the root to the vertex where the paths from the root to $x$ and
to $y$ bifurcate, and $p$-adic distance is therefore  
\begin{equation}\label{dperpnorm}
| x-y |_K =p^{-d_\perp(x,y)}, 
\end{equation}
see Figure~\ref{FigBTnorm}.

With this geometric picture in terms of Bruhat--Tits trees, we can reinterpret
the $p$-adic hierarchical lattices described in Definition~\ref{NumFlocmod}. 

In the Bruhat-Tits tree, at each vertex, one can identify the directions (edges
incident to that vertex) with a copy of $\P^1(\F_q)$, for $\F_q$ the residue field.
Given a root vertex $v_0$ in the tree (determined by a choice of a projective coordinate
in $\P^1(\K_\nu)$, as the meeting points of the geodesics connecting the points $\{ 0, 1, \infty \}$),
we refer to the subtrees stemming from the vertices adjacent to $v_0$ (oriented away from the root) 
as the {\em sectors} in the directions labelled by elements of $\P^1(\F_q)$.

\begin{lem}\label{BThierlocal}
Let $\K$ be a number field and $\nu\in \cV_{\K,{\rm fin}}$ a non-Archimedean place determined by a prime $\wp$ of $\K$.
The sets $\Sigma_{\K,\wp}$, $\bar\Sigma_{\K,\wp}$, $\cO_\K$, and 
$\cH_{\K,\wp}=\bar\Sigma_{\K,\wp}\cap \cO_\K$ are subsets of the boundary of
the sector of the Bruhat--Tits tree $\cT_{\K_\nu}$ that lies in the directions in
$\F_q=\P^1(\F_q)\smallsetminus \{ \infty \}$ from the root vertex. The function $J_{\alpha,\nu}(x-y)$
of \eqref{JNumFlocmod} extends to a function on the boundary $\partial\cT_{\K_\nu}$ given by
\begin{equation}\label{JtreeBT}
J_{\alpha,\nu}(x-y) = p^{- (1+\alpha) d_\perp(x,y)} \, , \ \  \  \forall x\neq y \in \P^1(K_\nu)=\partial\cT_{\K_\nu}\, , 
\end{equation}
where $d_\perp(x,y)$ is the number
of edges in the tree along the geodesic path in $\cT_{K_\nu}$ from the root vertex to the geodesic path between $x$ and $y$.
\end{lem}

\proof Consider the sector $\cS_{\K_\nu} \subset \cT_{\K_\nu}$ of the Bruhat--Tits tree $\cT_{\K_\nu}$ that lies in the directions in
$\F_q=\P^1(\F_q)\smallsetminus \{ \infty \}$ from the root vertex. This is referred to in the statistical physics literature
as the Bethe tree, or Bethe lattice. 
It has boundary $\partial \cS_{\K_\nu}=\cO_{\K_\nu} \subset \P^1(\K_\nu)=\partial \cT_{\K_\nu}$. 
Thus, $\cO_\K \subset \cO_{\K_\nu}$ is realized as a subset of the boundary of this sector. The sets $\Sigma_{\K,\wp}$
and $\bar\Sigma_{\K,\wp}$ of terminating and eventually periodic $\wp$-adic expansions, can be also identified as
sets of points in the boundary $\cO_{\K_\nu}=\partial \cS_{\K_\nu}$, where the coefficients of the $\wp$-expansion in
a set of representatives of $\cO_\K/\wp=\F_q$ provide the path in $\cT_{\K_\nu}$ from the root vertex to the boundary point.
The identification of \eqref{JNumFlocmod} and a restriction to this boundary region of \eqref{JtreeBT} follows from
\eqref{dperpnorm}. 
\endproof

\smallskip

The interpretation in terms of paths on Bruhat--Tits trees of the local percolation models 
at the non-Archimedean places of a number field, or at the finite places of a function field,
suggest a possible relation to the well studied percolation models on trees, \cite{AinzNew}, \cite{Lyons}.
Indeed, one can reinterpret the set $\Sigma_{\K,\wp}$ of terminating $\wp$-series as labeling
the internal vertices of the Bethe subtree $\cS_{\K_\nu} \subset \cT_{\K_\nu}$ of the Bruhat--Tits tree,
and the function $J_{\alpha,\nu}(x-y)$ can then be used to obtain a long-range percolation model
on $\cS_{\K_\nu}$. However, unlike in the usual percolation models on trees, here the function
$J_{\alpha,\nu}$ does not satisfy integrability, resulting in a different behavior. 

\begin{prop}\label{infclusterPadic}
The percolation model $\cP_\nu(\cH_\K)$ of Definition~\ref{NumFlocmod}, for $\nu\in \cV_{\K,{\rm fin}}$
a non-Archimedean place of a number field $\K$, has a non-integrable $J_{\alpha,\nu}$, and always
has an infinite cluster, with critical inverse temperature $\beta_{c,\nu}=0$.
\end{prop}

\proof The argument is analogous to Lemma~\ref{JqxInfSum} and Proposition~\ref{infclusterx}
for the equivalent systems for function fields, but we reformulate it here in terms of the
geometry of the Bruhat--Tits tree. Fix an element $x\in \cH_\K$, identified with a point in
the bounday $\partial \cS_{\K_\nu}=\cO_{\K_\nu}$ of the Bethe subtree of the Bruhat--Tits 
tree $\cT_{\K_\nu}$. Also fix an $N\in \N$. The level set 
$$ \cL_{\nu,N,x} := \{ y\in \cH_\K \,|\, J_{\alpha,\nu}(x-y) = q^{-(1+\alpha) N} \} $$
can be described geometrically in the following way. Consider the geodesic $\{ v_0, x \}$ 
in the Bruhat--Tits tree $\cT_{\K_\nu}$ from the root vertex $v_0$ to the boundary point $x$.
The condition that $v_\nu(x-y)=d_\perp(x,y)=N$ means that the path consisting of the
first $N$ edges of $\{ v_0, x \}$ starting at $v_0$ will end at a vertex $v_N$ of $\cT_{\K_\nu}$
which is on the geodesic $\{ x, y \}$ and is the first vertex of $\{ v_0, x \}$  that meets
this geodesic. Let $\cT_{\K_\nu,v_N}$ denote the subtree of $\cT_{\K_\nu}$ with root
at $v_N$ and consisting of all vertices below $v_N$ in the orientation away from $v_0$. 
It is then clear that $\partial \cT_{\K_\nu,v_N}\subseteq \cL_{\nu,N,x}$, hence $\# \cL_{\nu,N,x}=\infty$
hence $J_{\alpha,\nu}$ is non-integrable. 

The existence of an infinite cluster for any value of $\beta>0$ can be seen similarly.
 Fix an element $x\in \cH_\K \subset \partial \cS_{\K_\nu}$ and a subtree 
 $\cT_{\K_\nu,v}$ of $\cT_{\K_\nu}$ such that $x\notin \partial \cT_{\K_\nu,v}$
 and such that $\Omega_v = \cH_\K \cap \cT_{\K_\nu,v}$ is a countable set. 
 Such a set always exists: for example, consider the path $\{ v_0, x \}$ and 
 choose a number of steps $N$ along this path. Let  $a\in \F_q$ be the next direction at $v_N$ of
 the next edge of the path $\{ v_0, x \}$, let $b\neq a \in \F_q$ be a different
 direction at $v_N$ and let $v$ be the other endpoint of the edge in the
 $b$ direction. Then the set $\Omega_v = \cH_\K \cap \cT_{\K_\nu,v}$ has
 the desired properties. We then see that the probability of having no edges
 between $x$ and $\Omega_v$ is given by
 $$ \prod_{y\in \Omega_v} e^{-\beta J_{\alpha,\nu}(x-y)} =e^{-\beta \sum_{y\in \Omega_v} J_{\alpha,\nu}(x-y)} \, . $$
 Thus, it suffices to show that $\sum_{y \in \Omega_v} J_{\alpha,\nu}(x-y)$ is divergent, which
 now follows from the argument above, as $d_\perp(x,y)=N$ for all $y\in \Omega_v$ by
 construction. 
 This shows that there are some edges between $x$ and $\Omega_v$. Choose a $y\in \Omega_v$
 connected to $x$ by an edge, and consider the complement $\Omega_v\smallsetminus \{ y \}$
 and a region $\Omega_w\subset \Omega_v\smallsetminus \{ y \}$ constructed in the following way.
 Consider the geodesic $\{ v, y \}$ in $\cT_{\K_\nu}$, and let $a_v\in \F_q$ be the direction at $v$
 of the path $\{ v, y \}$. Choose a $b\neq a_v \in \F_q$ and let $w$ be the other endpoint of
 the edge at $v$ in the direction $b$. Let $\Omega_w=\Omega_v\cap \cT_{\K_\nu, w}$. The same
 argument then applies to $y$ and $\Omega_w$, showing there are edges connecting them.
 Iterating this argument shows the existence of an infinite cluster. 
\endproof

Note that, while the same argument applies to each $\cP_\nu(\cH_\K)$, each of these
graphs now lives in a different $\cT_{\K_\nu,v}$. Thus, the existence of an infinite
cluster in each $\cP_\nu(\cH_\K)$ does not imply that the {\em same} infinite cluster
would exist at all $\nu\in \cV_{\K,{\rm fin}}$, as already observed in the function fields case.
Indeed, the difference is what makes the adelic case
much more interesting that the individual local non-Archimedean cases, and directly 
related to long-range percolation on lattices, as we discuss in \S \ref{NumFadelicSec}.
We first need to discuss the corresponding local percolation model at the Archimedean
places, so that the adelic formula can be applied.

\subsection{Percolation model at the Archimedean places} \label{ArchimPlSec}

For a number field $\K$ with $[\K,\Q]=n$, the Minkowski embedding 
\begin{equation}\label{Mink}
\sigma_M: \K \hookrightarrow 
\K_\R=\R \otimes_\Q \K \simeq \R^n 
\end{equation}
is given
by $x\mapsto (\sigma_1(x),\ldots, \sigma_r(x), \Re \sigma_{r+1}(x), \Im \sigma_{r+1}(x),
\ldots \Re \sigma_{r+s}(x), \Im \sigma_{r+s}(x))$, for $\{ \sigma_i \}_{i=1}^r$ the set of real
embeddings of $\K$ and $\{ \sigma_i, \bar\sigma_i \}_{i=r+1}^{r+s}$ the set of 
complex embeddings (which come in conjugate pairs), which together give all the
$n=r+2s$ archimedean places of $\K$.

\smallskip

\begin{lem}\label{numKtransLat}
Let $\K$ be a number field with $[\K:\Q]=n$. Its ring of integers $\cO_\K$
is a $(r,s)$-mixed transverse lattice (in the sense of Definition~\ref{transLatmix}) 
in the Euclidean space $\K_\R=\R \otimes_\Q \K \cong\R^r\times \C^s\cong \R^n$,
under the Minkowski embedding.  In the case where $\K$ is a totally real field, the Minkowski
embedding of $\cO_\K$ is a transverse lattice in the sense of Definition~\ref{transLat}. 
\end{lem}

\proof For every $\sigma\in \cV_{\K,\infty}$, the set of infinite, archimedean, places of $\K$, 
we have an embedding $\sigma: \K \hookrightarrow \C$, of which $r_1$ are
real embeddings and $r_2$ are number of complex conjugate pairs of  
complex embeddings, with $n=r_1 + 2 r_2$. Let $|\cdot |_\alpha$ be the associated
Euclidean norm in $\R$ or $\C$. Since $\sigma$ is an embedding of $\K$, for $x\in \cO_\K$
we have $| x |_\sigma =0$ iff $x=0$. Thus, the image 
$\Lambda=\{ (\sigma(x))_{\sigma\in \cV_{\K,\infty}}\in \R^n \,|\, x\in \cO_\K \}$ 
is a lattice in $\R^n$ with $\Lambda\cap \cD_n(\R)=\{ 0 \}$. 
\endproof

\smallskip

\begin{defn}\label{archimPHKmod}{\rm
Let $\K$ be a number field and let $\sigma \in \cV_{\K,\infty}$ be an Archimedean place of $\K$,
namely a (real or complex) embedding $\sigma: \K \hookrightarrow \C$.  The local long-range percolation
model $\cP_\sigma(\cH_\K)$ is the random graph with set of vertices $\cH_\K$ and with inclusion
probabilities
\begin{equation}\label{ProbsigmaHK}
\P_{\beta,\alpha,\sigma}(\{ x, y \}) = 1 - e^{-\beta J_{\alpha,\sigma}(x-y)} \, , 
\end{equation}
where
\begin{equation}\label{JsigmaHK}
J_{\alpha,\sigma}(x-y):= | x-y |_\sigma^{-(1+\alpha)} = | \sigma(x)-\sigma(y) |_{\K_\sigma}^{-(1+\alpha) n_\sigma} \, ,
\end{equation}
where $\K_\sigma=\R$ or $\K_\sigma=\C$ depending on whether $\sigma$ is a real or complex embedding,
and $|\cdot |_{\K_\sigma}$ is the absolute value on $\R$ or $\C$, where $n_\sigma=1$ for $\K_\sigma=\R$
and $n_\sigma=2$ for $\K_\sigma=\C$. 
}\end{defn}

\smallskip

We can also assemble these Archimedean local models over the infinite places of $\K$. The
resulting model recovers the toric percolation model of Definition~\ref{toricpercLat} 
on the lattice $\cO_\K$.

\begin{defn}\label{infadelesPHKmod}{\rm
Let $\K$ be a number field with $[\K:\Q]=n$ and $\cV_{\K,\infty}$ the set of the $n$ Archimedean places of $\K$. 
The Archimedan adelic long-range percolation
model $\cP_{\A_{\K,\infty}}(\cH_\K)$ is the random graph with set of vertices $\cH_\K$ and with inclusion
probabilities
\begin{equation}\label{ProbsigmaHK}
\P_{\beta,\alpha,\A_{\K,\infty}}(\{ x, y \}) = \prod_{\sigma\in \cV_{\K,\infty}} \P_{\beta,\alpha,\sigma}(\{ x, y \}) \, , 
\end{equation}
with $\P_{\beta,\alpha,\sigma}$ as in \eqref{ProbsigmaHK}, \eqref{JsigmaHK}.
}\end{defn}

\begin{prop}\label{toricAinfty}
For $x,y\in \cH_\K \subset \cO_\K$ with large $|\sigma(x)-\sigma(y)|_{\K_\sigma}$, for $\sigma\in \cV_{\K,\infty}$, the probability
of the Archimedan adelic long-range percolation model $\cP_{\A_{\K,\infty}}(\cH_\K)$ behaves like the probability of the
long-range toric percolation model of Definition~\ref{toricpercLat}, on the lattice $\cO_\K$ in the Minkowski embedding 
$\cO_\K \hookrightarrow \K_\R$,
\begin{equation}\label{toricAKinfP}
\P_{\beta,\alpha/n,\A_{\K,\infty}}(\{ x, y \}) \sim  \P_{\beta^n,\alpha,\bT(r,s)}(\{ (\sigma_\R(x),\sigma_\C(x)),(\sigma_\R(y),\sigma_\C(y)) \})\, ,
\end{equation}
with $\P_{\beta,\alpha,\bT(r,s)}$ as in \eqref{mixPtoric}, with $\sigma_\R(x)=(\sigma_1(x),\ldots,\sigma_r(x))$ the
real embeddings and $\sigma_\C(x)=(\sigma_{r+1}(x),\ldots, \sigma_{r+s}(x))$
the complex embeddings.
\end{prop} 

\proof For large $|\sigma(x)-\sigma(y)|_{\K_\sigma}$, $\sigma\in \cV_{\K,\infty}$, that is, for small  $J_{\alpha,\sigma}(x-y)$, we have
$$ \P_{\beta,\alpha/n,\A_{\K,\infty}}(\{ x, y \}) \sim \beta^n \prod_{\sigma \in \cV_{\K,\infty}} J_{\alpha/n,\sigma}(x-y) =
 \beta^n \prod_{\sigma \in \cV_{\K,\infty}} |\sigma(x)-\sigma(y)|_{\K_\sigma}^{-(1+\alpha/n)n_\sigma} $$ $$ =
 \beta^n J_{\alpha,0,1/n}(\sigma_\R(x)-\sigma_\R(y),\sigma_\C(x)-\sigma_\C(y)) $$ $$ \sim 
 \P_{\beta^n,\alpha,\bT(r,s)}(\{ (\sigma_\R(x),\sigma_\C(x)),(\sigma_\R(y),\sigma_\C(y)) \}) \, , $$
with $J_{\alpha,0,1/N}$ as in \eqref{Jalpha0NM0}. 
\endproof 

\subsection{Number fields: adelic percolation model}  \label{NumFadelicSec}

We can then similarly assemble the local models at the non-Archimedean places, as in 
Definition~\ref{NumFlocmod} into a single model over the finite adeles $\A_{\K,{\rm fin}}$ of a number field $\K$ 
and compare the resulting model to the one associated to $\A_{\K,\infty}$ in Definition~\ref{infadelesPHKmod}. 

\begin{defn}\label{defNumFfinadelesPerc}{\rm Let $\K$ be a number field and let $\cV_{\K,{\rm fin}}$ be
the set of non-Archimedean places. The non-Archimedan adelic long-range percolation
model $\cP_{\A_{\K,{\rm fin}}}(\cH_\K)$ is the random graph with set of vertices $\cH_\K$ and with inclusion
probabilities
\begin{equation}\label{ProbsigmaHK}
\P_{\underline{\beta},\underline{\alpha},\A_{\K,{\rm fin}}}(\{ x, y \}) 
= \prod_{\nu \in \cV_{\K,{\rm fin}}} \P_{\beta_\nu,\alpha_\nu,\nu}(\{ x, y \})\, ,
\end{equation}
with $\P_{\beta_\nu,\alpha,\nu}$ as in \eqref{ProbNumFlocmod}, and 
with $\underline{\beta}=(\beta_\nu)_{\nu \in \cV_{\K,{\rm fin}}}$ and
$\underline{\alpha}=(\alpha_\nu)_{\nu \in \cV_{\K,{\rm fin}}}$.
}\end{defn}

\smallskip

As in the case of function fields, we make a choice of the sequences
$\underline{\beta}=(\beta_\nu)_{\nu \in \cV_{\K,{\rm fin}}}$ and
$\underline{\alpha}=(\alpha_\nu)_{\nu \in \cV_{\K,{\rm fin}}}$
that is dictated by the underlying geometry.

\begin{defn}\label{betanuZetaDef}{\rm
For a number field $\K$ we take $\underline{\beta}=(\beta_\nu)_{\nu \in \cV_{\K,{\rm fin}}}$ to be of the form
\begin{equation}\label{betanuZeta}
\beta_\nu = \beta \cdot f_{\K_\nu} \, ,
\end{equation}
where $f_{\K_\nu}=[\F_q: \F_p]$ for $\F_q=\cO_{\K_\nu}/\fm_{\K_\nu}$, and $p$ the characteristic of the residue field.
Correspondingly, we take 
$\underline{\alpha}=(\alpha_\nu)_{\nu \in \cV_{\K,{\rm fin}}}$ to be either constant, $\alpha_\nu=\alpha$
for all $\nu \in \cV_{\K,{\rm fin}}$, or else of the form
\begin{equation}\label{alphanuZeta}
\alpha_\nu = \alpha + \log_p f_{\K_\nu}\, .
\end{equation}
}\end{defn}

The Dedekind zeta function of a number field $\K$ has an Euler product expression of the form
\begin{equation}\label{zetaKDed}
\zeta_\K(s) = \prod_\wp (1 - N(\wp)^{-s})^{-1} \, ,
\end{equation}
with $\wp$ ranging over the non-zero prime ideals of $\cO_\K$. In the case of $\K=\Q$ this is the
Euler product expansion of the Riemann zeta function $\zeta(s)=\prod_p (1-p^{-s})^{-1}$. The
Euler product converges absolutely for $\Re(s)>1$, and $\zeta_\K(s)$ has analytic continuation
to a meromorphic function in the complex plane, with a single simple pole at $s=1$.
For $S$ a finite set of prime ideals of $\cO_\K$, we also write
\begin{equation}\label{zetaKDed}
\zeta^{(S)}_\K(s) = \prod_{\wp \in S^c} (1 - N(\wp)^{-s})^{-1} \, ,
\end{equation}
for the Dedekind zeta function with those Euler factors removed. In the analytic
continuation region of $\zeta_\K(s)$, \eqref{zetaKDed} has continuation of
the form $\zeta^{(S)}_\K(s)=\zeta_\K(s)  \cdot \prod_{\wp\in S} (1 - N(\wp)^{-s})$.

\smallskip

\begin{rem}\label{zeroDed}{\rm 
Unlike the case of function fields, and with the exception of 
the case of $\K=\Q$, where the Riemann
zeta function does not have any zeros on the positive real line, 
the Dedekind zeta function of a number field $\K$ can have a 
single simple zero on the positive real line, located close to $s=1$ (Siegel zero). }
\end{rem}

\smallskip

For a number field $\K$, let $d_\K$ denote the absolute value of the
discriminant. There are estimates that restrict the location of the Siegel 
zero of the Dedekind zeta function to a subinterval $\beta_0< \beta < 1$, 
with $\beta_0$ a function of $d_\K$. For example, the result 
of \cite{Stark} shows that, for an arbitrary number field, one can take
\begin{equation}\label{beta0Stark}
\beta_0 = 1-\frac{1}{4 \log d_\K} \, ,
\end{equation}
while for sufficiently large $d_\K$, one can take
\begin{equation}\label{beta0Kad}
\beta_0 =1-\frac{1}{12.74 \, \log d_\K}
\end{equation}
as shown in \cite{Kadiri}.

\smallskip

The long range toric percolation model
$\cP_\bT(\cO_\K)$ on the lattice $\cO_\K$ with the Minkowski embedding 
$\cO_\K \hookrightarrow \K_\R$, restricted to $\cH_\K \subset \cO_\K$, 
behaves like the non-Archimedan adelic long-range percolation
model $\cP_{\A_{\K,{\rm fin}}}(\cH_\K)$, in the following sense.

\smallskip

\begin{thm}\label{NumFadelicfininftyPerc}
Let $\K$ be a number field with $[\K:\Q]=n$, and $\cV_\K=\cV_{\K,{\rm fin}}\cup \cV_{\K, \infty}$ its set of places. 
Let $d_\K$ be the absolute value of the discriminant of $\K$. Let 
$\underline{\beta}=(\beta_\nu)_{\nu \in \cV_{\K,{\rm fin}}}$ be as in \eqref{betanuZeta}. Let $S\subset 
\cV_{\K,{\rm fin}}$ be a finite (large) set of non-Archimedean places of $\K$.  Let
\begin{equation}\label{betaASnumF}
\beta_{\A,S}^n=\left\{ \begin{array}{ll}  \frac{\beta^{\# S}}{\zeta_\K(\beta)} & \text{for } \beta\leq 1 \\[3mm]
 \frac{\beta}{\zeta_\K(\beta)} & \text{for } \beta\leq 1
\end{array}\right.  \ \ \ \text{ and } \ \ \ 
{\beta^{\prime\, n}_{\A,S}}=\left\{ \begin{array}{ll}  \frac{\beta}{\zeta^{(S)}_\K(\beta)} & \text{for }  \beta\leq 1 \\[3mm]
\frac{\beta^{\# S}}{\zeta^{(S)}_\K(\beta)} & \text{for } \beta > 1 \, . 
\end{array}\right.
\end{equation}
Suppose given $x,y \in \cH_\K$ with large 
$|\sigma(x)-\sigma(y)|_{\K_\sigma}$, for $\sigma\in \cV_{\K,\infty}$, and such that $S_{x-y}\subset S$,
for $S_{x-y}\subset \cV_{\K,{\rm fin}}$ the set of non-archimedean places with valuation $v_\nu(x-y)\neq 0$.
Let $\beta_0$ be as in \eqref{beta0Stark}, or (for large $d_\K$) as in \eqref{beta0Kad}.
\begin{enumerate}
\item For any $\beta \geq 1$ or $0<\beta\leq \beta_0<1$ and 
for $\underline{\alpha}$ with constant $\alpha_\nu =\alpha$ 
\begin{equation}\label{toricAdelicnonArch}
\P_{\beta_{\A,S}^n, \alpha, \bT}(\{ x, y \}) \leq \P_{\underline{\beta},\alpha/n,\A_{\K,{\rm fin}}}(\{ x, y \}) \, ,
\end{equation}
\item In the same range of $\beta$ and 
for $\underline{\alpha}$ with $\alpha_\nu =\alpha + \log_p f_{\K_\nu}$ as in \eqref{alphanuZeta}
\begin{equation}\label{toricAdelicnonArch2}
\P_{\underline{\beta},\underline{\alpha}/n,\A_{\K,{\rm fin}}}(\{ x, y \}) \leq \P_{{\beta^{\prime\, n}_{\A,S}}, \alpha, \bT}(\{ x, y \})  \, . 
\end{equation}
\end{enumerate}
\end{thm}

\proof  For a place $\nu\in \cV_{\K,{\rm fin}}$ of a number field $\K$ corresponding to a prime $\wp$ of $\K$ 
over a rational prime $p$, we have
$\F_q=\cO_{\K_\nu}/\fm_{\K_\nu}=\cO_\K/\wp$, hence $N(\wp)=p^{[\cO_\K/\wp:\F_p]}=p^{f_{\K_\nu}}$.
Thus, we can write \eqref{betanuZeta} as 
$$ \beta_\nu = \beta \cdot \log_p N(\wp) \, . $$
We have, for large $| x-y |_\nu$ at $\nu \in S_{x-y}$,
$$ \P_{\underline{\beta},\underline{\alpha},\A_{\K,{\rm fin}}}(\{ x, y \}) \sim \prod_{\nu \in S_{x-y}^c} (1- e^{-\beta_\nu}) \cdot
\prod_{\nu \in S_{x-y}} \beta_\nu \, | x-y |_\nu^{(1+\alpha_\nu)} $$ $$ =
 \prod_{\wp \in S_{x-y}^c} (1- N(\wp)^{-\beta}) \cdot \beta^{\# S_{x-y}} \cdot \prod_{\nu \in S_{x-y}} f_{\K_\nu} \cdot
 \prod_\nu | x-y |_\nu^{(1+\alpha_\nu)} \, . $$
In the first case, arguing as in Lemma~\ref{betaZetaProd}, 
we obtain, for $\alpha_x=\alpha$ constant, and for large $| x-y |_\nu$, for $\nu \in S_{x-y}$,
$$ \P_{\underline{\beta},\alpha,\A_{\K,{\rm fin}}}(\{ x, y \}) \sim \zeta^{(S_{x-y})}_\K(\beta)^{-1} \cdot D_{S_{x-y}} \cdot
\beta^{\# S_{x-y}} \cdot \prod_{\nu \in S_{x-y}}  | x-y |_{\nu}^{(1+\alpha)}\, ,  $$
where
$$ D_{S_{x-y}} = \prod_{\nu \in S_{x-y}} f_{\K_\nu} \, . $$
In the second case, as in  Lemma~\ref{betaZetaProd},  
for $\alpha_x=\alpha+ \log_p f_{\K_\nu}$, we have, in the same range,
$$ \P_{\underline{\beta},\underline{\alpha},\A_{\K,{\rm fin}}}(\{ x, y \}) \sim \zeta^{(S_{x-y})}_\K(\beta)^{-1} \cdot \tilde D_{S_{x-y}} \cdot \beta^{\# S_{x-y}} \cdot \prod_{\nu \in S_{x-y}}  | x-y |_{\nu}^{(1+\alpha)}\, ,  $$
where
$$ \tilde D_{S_{x-y}} = \prod_{\nu \in S_{x-y}} f_{\K_\nu}^{1-v_\nu(x-y)} \, . $$
By Proposition~\ref{toricAinfty} we also know that we have
$$ \P_{\beta^n, \alpha, \bT}(\{ x, y \}) \sim \P_{\beta,\alpha/n,\A_{\K,\infty}}(\{ x, y \})\, . $$
where, for large $|\sigma(x)-\sigma(y)|_{\K_\sigma}$, at the Archimedean places, we have
$$ \P_{\beta,\alpha/n,\A_{\K,\infty}}(\{ x, y \}) \sim \beta^n \prod_{\sigma\in \cV_{\K,\infty} }
| x-y |_\sigma^{-(1+\alpha/n)} \, . $$
The adelic formula for the norm of elements of $\K\smallsetminus \{ 0 \}$ then gives
$$ \prod_{\sigma \in \cV_{\K,\infty} } | x-y |_\sigma^{-(1+\alpha/n)} = \prod_{\nu\in \cV_{\K,{\rm fin}} }
| x-y |_\nu^{(1+\alpha/n)} \, . $$

Thus, combining these relations we obtain that, in this range with large 
$|\sigma(x)-\sigma(y)|_{\K_\sigma}$, for $\sigma\in \cV_{\K,\infty}$, we
have the following cases. In the first case of constant $\alpha_\nu=\alpha$,
$$ \P_{\underline{\beta},\underline{\alpha}/n,\A_{\K,{\rm fin}}}(\{ x, y \}) \sim
\zeta^{(S_{x-y})}_\K(\beta)^{-1} \cdot D_{S_{x-y}} \cdot
\beta^{\# S_{x-y}} \cdot \prod_{\sigma\in \cV_{\K,\infty}}  | x-y |_{\sigma}^{-(1+\alpha/n)} $$
$$ \geq \zeta_\K(\beta)^{-1} \beta^M \prod_{\sigma\in \cV_{\K,\infty}}  | x-y |_{\sigma}^{-(1+\alpha/n)} 
= \beta_{\A,S}^n  \prod_{\sigma\in \cV_{\K,\infty}}  | x-y |_{\sigma}^{-(1+\alpha/n)} $$
$$ \sim \P_{\beta_{\A,S},\alpha/n,\A_{\K,\infty}}(\{ x, y \}) \sim  \P_{\beta_{\A,S}^n, \alpha, \bT}(\{ x, y \}) \, , 
$$
where $M=1$ if $\beta > 1$ and $M=\# S$ if $\beta\leq  1$. Similarly, in the second case with 
$\alpha_\nu=\alpha + \log_p f_{\K_\nu}$ we have
$$ \P_{\underline{\beta},\underline{\alpha}/n,\A_{\K,{\rm fin}}}(\{ x, y \}) \sim
\zeta^{(S_{x-y})}_\K(\beta)^{-1} \cdot \tilde D_{S_{x-y}} \cdot
\beta^{\# S_{x-y}} \cdot \prod_{\sigma\in \cV_{\K,\infty}}  | x-y |_{\sigma}^{-(1+\alpha/n)} $$
$$ \leq \zeta^{(S)}_\K(\beta)^{-1} \beta^M \prod_{\sigma\in \cV_{\K,\infty}}  | x-y |_{\sigma}^{-(1+\alpha/n)} 
= {\beta^{\prime\, n}_{\A,S}} \, \, \prod_{\sigma\in \cV_{\K,\infty}}  | x-y |_{\sigma}^{-(1+\alpha/n)} $$
$$ \sim \P_{\beta'_{\A,S},\alpha/n,\A_{\K,\infty}}(\{ x, y \}) \sim  \P_{{\beta^{\prime\, n}_{\A,S}}, \alpha, \bT}(\{ x, y \})\, ,  $$
where in this case $M=1$ if $\beta\leq  1$ and $M=\# S$ if  $\beta > 1$. 
Thus, we obtain \eqref{toricAdelicnonArch} and \eqref{toricAdelicnonArch2}.

As observed in Remark~\ref{zeroDed},  in the case of number fields, the 
expressions for $\beta_{\A,S}$ and $\beta'_{\A,S}$ in \eqref{betaASnumF}, as functions
of $\beta >0$ can have a singular point, for a $\beta$ somewhere in the interval $0< \beta <1$,
due to the possible presence of a Siegel zero of the Dedekind zeta function.
Thus, this restricts our estimates to a part of the interval that is free of zeros. We
can take $0< \beta \leq \beta_0$, with $\beta_0$ as in \eqref{beta0Stark} (or as in \eqref{beta0Kad}, 
when $d_\K$ is sufficiently large; see \cite{Kadiri} for more details).   
\endproof

\smallskip

As in the case of function fields, we can use the result of Theorem~\ref{NumFadelicfininftyPerc} to
obtain estimates on the critical temperature for the adelic system. 

\begin{prop}\label{betacToric}
Let $\underline{\beta}=(\beta_\nu)_{\nu\in \cV_{\K,{\rm fin}}}$ for a number field $\K$, with $\beta_\nu =\beta \, f_{\K_\nu}$
and let $\underline{\alpha}=(\alpha_\nu)_{\nu\in \cV_{\K,{\rm fin}}}$ be either constant $\alpha_\nu=\alpha$ or of the form
$\alpha_\nu=\alpha+\log_p f_{\K_\nu}$. Let $\beta_{c,\bT}$ be the critical inverse temperature of the toric percolation
model on the lattice $\cH_\K \subset \cO_\K$. Then, for $\beta$ satisfying 
$$ \beta > \max\{ 1, \beta_{c,\bT} \cdot \zeta_\K(\beta) \} $$
there is an infinite cluster with non-zero probability in the adelic model with 
inclusion probabilities $\P_{\underline{\beta},\alpha, \A_{\K,{\rm fin}}}(\{x, y\})$ with constant $\alpha_\nu=\alpha$.
For $\beta$ satisfying
$$ \beta \leq \min \{ \beta_0, \beta_{c,\bT} \cdot \zeta_\K(\beta) \} \, , $$
with $\beta_0$ as in Theorem~\ref{NumFadelicfininftyPerc}, there is no infinite cluster with
non-zero probability in the adelic model with inclusion probabilities $\P_{\underline{\beta},\underline{\alpha}, 
\A_{\K,{\rm fin}}}(\{x, y\})$ with $\alpha_\nu =\alpha + \log_p f_{\K_\nu}$.
\end{prop}

\proof The result follows from Theorem~\ref{NumFadelicfininftyPerc} in the same way as in
Proposition~\ref{betaAScritical} for the case of function fields.
\endproof

\subsection{The adelic relation between hierachical and ordinary lattice percolation} \label{FinSec}

We say that two percolation models $\cP$ and $\cP'$ are {\em equivalent} if the respective
percolation probabilities can be matched under an appropriate mapping of the underlying
spaces. We say that they can be {\em compared} if there are specific fine tunings of parameters in both models 
that yield estimates in both directions relating the respective percolation probabilities.  

\medskip

We can now combine all the relations between the various percolation models described in the previous sections and 
obtain the following statement that makes the content of the diagram \eqref{bigdiagram} more precise.

\begin{thm}\label{thmdiagr}
Let $\K$ be a number field with $[\K:\Q]=n$ and such that $\cH_\K \subset \cO_\K$ is a countable subset.
Then the relation between the lattice percolation model $\cP_{\rm latt}(\cO_\K)$ on the ring
of integers  in the Minkowski embedding $\cO_\K\subset \R^n$ is related to the long range percolation
models on hierarchical lattices through adelic models and the power mean long range percolation model
in the following way, illustrated in diagram \eqref{bigdiagram},
\begin{enumerate}
\item The long range percolation model $\cP_{\rm hier}(\cH_q^1)$ on the hierarchical lattice $\cH^1_q$ is equivalent to
the long range percolation model $\cP_{\bA_{\F_q(C),\infty}}(\cH^1_q)$ at the point at infinity $\infty\in C$ of a function field $\F_q(C)$.
\item In the long range (large $\| f-g \|$) the behavior of the model $\cP_{\rm hier}(\cH_q^1)$ can be compared to the 
behavior of the adelic model $\cP_{\bA_{\F_q(C),{\rm fin}}}(\cH^1_q)$, for the data 
$\underline{\beta}=(\beta_x)_{x\in \cV_{\F_q(C),{\rm fin}}}$ with $\beta_x=\beta \deg(x)$ and $\underline{\alpha}=(\alpha_x)$
with either constant $\alpha_x=\alpha$ or $\alpha_x=\alpha+ \log_q \deg(x)$, for a subregion of the hierarchical lattice
depending on the choice of a finite set $S$ of places of $\F_q(C)$.
\item The function field model $\cP_{\bA_{\F_q(C),x\neq \infty}}(\cH_q^1)$, with $q=p^r$ 
at a finite points $x\in C$ has the same behavior as the number field model 
$\cP_{\bA_{\K,{\rm fin}}, \nu}(\cH_\K)$ at a non-archimedean place $\nu\in \cV_{\K,{\rm fin}}$ with residue
field of characteristic $p$. 
\item In the long range,  $\| x-y \|$ large and $x,y\in \cO_\K$ in a subregion of the lattice that depends on the choice of a 
finite set $S$ of non-Archimedean places of $\K$, the behavior of the adelic model $\cP_{\bA_{\K,{\rm fin}}}(\cH_\K)$, for
a number field $\K$ for $\underline{\beta}=(\beta_\nu)_{\nu\in \cV_{\K,{\rm fin}}}$ with $\beta_\nu =\beta\, f_{\K_\nu}$,
and $\underline{\alpha}=(\alpha_\nu)$
with $\alpha_\nu=\alpha$ (or with $\alpha_\nu =\alpha+ \log_p f_{\K_\nu}$)
can be compared with the behavior of the adelic model $\cP_{\bA_{\K,\infty}}(\cH_\K)$ with $\alpha/n$ and
inverse temperature $\beta_{\A,S}$ (respectively, $\beta'_{\A,S}$) that is a function of $\beta$ 
and $\K$ and $S$, in a range of values $\beta \geq 1$ and $0<\beta \leq \beta_0$ where 
$\beta_0$ is a function of the discriminant of the number field $d_\K$ that identifies a region
free of possible real zeros of the Dedekind zeta function.
\item The  model $\cP_{\bA_{\K,\infty}}(\cH_\K)$ with parameters $\beta$ and $\alpha/n$ is equivalent to the toric long 
range percolation model $\cP_\bT(\cH_\K)$ with parameters $\beta^n$ and $\alpha$.
\item The toric long range percolation model $\cP_\bT(\cO_\K)$ 
and the lattice long range percolation model 
$\cP_{\rm latt}(\cO_\K)$ on $\cO_\K$ in the Minkowski embedding, restricted to $\cH_\K$, are the special values
$t=0$, $\lambda=1/n$ and $t=2$, $\lambda=1/n$, respectively, of the power mean long range percolation 
model $\cP_{\cM_t}(\cO_\K)$. 
\end{enumerate}
\end{thm}

\proof Consider again the diagram of \eqref{bigdiagram}, which we now draw as
$$
\xymatrix{  \cP_{\bA_{\F_q(C),\infty}}(\cH^1_q) \ar[d]^{\textcircled{1}} \ar[r]^{\textcircled{2}}  & \cP_{\bA_{\F_q(C),{\rm fin}}}(\cH^1_q) \ar[l] \ar[d]^{\textcircled{r}}  & \cP_{\bA_{\K,{\rm fin}}}(\cH_\K) \ar[d]^{\textcircled{r}}   \ar[r]^{\textcircled{4}} & \cP_{\bA_{\K,\infty}}(\cH_\K) \ar[d]^{\textcircled{5}} \ar[l]    & \\
 \cP_{\rm hier}(\cH_q^1) \ar[u]  & \cP_{\bA_{\F_q(C),x\neq \infty}}(\cH_q^1) \ar[r]^{\textcircled{3}} & \cP_{\bA_{\K,{\rm fin}}, \nu}(\cH_\K)  \ar[l]  & \cP_\bT(\cH_\K) \ar[u] \ar[r]^{\textcircled{6}} & \cP_{\rm latt}(\cH_\K)  \ar[l] }
$$
where the numbered arrows correspond to the numbering of the statements. 
The arrow marked as $\textcircled{1}$ is the content of Proposition~\ref{abgrpHNLFqt}. The arrows marked 
as $\textcircled{2}$ and $\textcircled{4}$ are the result of the adelic product formula, respectively as shown in Lemma~\ref{lemJxJinfty}
and Theorem~\ref{localadelicFF} for the function field case, and in Theorem~\ref{NumFadelicfininftyPerc} for
number fields (where the possible presence of a Siegel zero of the Dedekind zeta function plays a role). 
The arrows marked $\textcircled{r}$ are, in both cases, the restriction 
to the component at one of the
finite places of the global field. 
The arrow $\textcircled{3}$ is obtained by comparing Proposition~\ref{infclusterPadic} with Lemma~\ref{JqxInfSum} and
Proposition~\ref{infclusterx}, or by observing that both models have an analogous geometric description in terms of
Bruhat--Tits trees of the respective local fields. 
The arrow marked $\textcircled{5}$ is the equivalence shown in Proposition~\ref{toricAinfty}.
The arrow $\textcircled{6}$ is the result of Lemma~\ref{toricPMperc} and Remark~\ref{t2Mt}.
The hypotheses of the statement correspond to the hypotheses of each of these individual statements that establish the relations
illustrated in the arrows of the diagram.
\endproof

About the arrow $\textcircled{3}$ above, note that while local 
models $\cP_{\bA_{\F_q(C),x\neq \infty}}(\cH_q^1)$ and $\cP_{\bA_{\K,{\rm fin}}, \nu}(\cH_\K)$
with residue fields of the same characteristics have a similar formulation on the corresponding Bruhat--Tits trees, the
adelic products of such local models look different. In the number field case of $\cP_{\bA_{\K,{\rm fin}}, \nu}(\cH_\K)$
with non-Archimedean places $\nu \in \cV_{\K,{\rm fin}}$ determined by primes $\wp$ of $\K$, the characteristics of
the residue field ranges over all the primes $p$, hence the branching structure of the Bruhat--Tits tree correspondingly varies, 
while in the function fields case of $\cP_{\bA_{\F_q(C),{\rm fin}}}(\cH_q^1)$ the cardinality remains a single fixed $p$, 
and the branching is always just by powers $q=p^r$. Note also that the individual systems 
$\cP_{\bA_{\F_q(C),x\neq \infty}}(\cH_q^1)$ and $\cP_{\bA_{\K,{\rm fin}}, \nu}(\cH_\K)$ are 
insensitive to different choices of $\beta_x$, $\beta_\nu$, because of the non-integrability condition of
Proposition~\ref{infclusterPadic} and Lemma~\ref{JqxInfSum}, but the corresponding adelic systems
are highly sensitive to the data $\underline{\beta}$ and $\underline{\alpha}$. In particular, the behavior 
of the adelic system will be different for $\cP_{\bA_{\K,{\rm fin}}}(\cH_\K)$ and $\cP_{\bA_{\F_q(C),{\rm fin}}}(\cH_q^1)$
as the data $\underline{\beta}$ and $\underline{\alpha}$ reflect the different underlying arithmetic
geometry of function fields and number fields.

\subsection*{Acknowledgment}
This work is supported by NSF grant DMS-2104330. I am grateful to Tom Hutchcroft for very
helpful discussions and for first explaining to me the percolation models on lattices and hierarchical lattices 
and their intriguing relations. I thank Gunther Cornelissen for providing useful comments and feedback 
on a draft of this manuscript, and Yassine El-Maazouz for helpful conversations. I thank the anonymous referee
for a very careful reading and for many useful comments. 

\subsection*{Conflict of Interest and Data Availability statements}

The author declares no conflict of interests. There are no data associated to this work.

\end{document}